\documentclass[fleqn,usenatbib]{mnras}

\usepackage{newtxtext,newtxmath}

\usepackage[T1]{fontenc}

\DeclareRobustCommand{\VAN}[3]{#2}
\let\VANthebibliography\thebibliography
\def\thebibliography{\DeclareRobustCommand{\VAN}[3]{##3}\VANthebibliography}

\usepackage{graphicx}	
\usepackage{amsmath}
\usepackage{multicol}

\usepackage{subcaption}

\title[Flux density monitoring]{Flux density monitoring of 89 millisecond pulsars with MeerKAT}

\author[P. Gitika et al.]{
P. Gitika$^{1,2}$,\thanks{E-mail: pgitika@swin.edu.au}
M. Bailes$^{1,2}$,
R. M. Shannon$^{1,2}$,
D. J. Reardon$^{1,2}$,
A. D. Cameron$^{1,2}$,
M. Shamohammadi$^{1,2}$,\newauthor
M. T. Miles$^{1,2}$,
C. M. L. Flynn$^{1,2}$,
A. Corongiu$^{3}$,
M. Kramer $^{4}$
\\
$^{1}$ Centre for Astrophysics and Supercomputing, Swinburne University of Technology, Mail H39, PO Box 218, VIC 3122, Australia.\\
$^{2}$ARC Centre of Excellence for Gravitational Wave Discovery (OzGrav), Swinburne University of Technology, Mail H11, PO Box 218, VIC 3122.\\
$^{3}$INAF - Osservatorio Astronomico di Cagliari, Via della Scienza 5, I-09047 Selargius.\\
$^{4}$MPI f\"ur Radioastronomie, Auf dem H\"ugel 69, 53121 Bonn, Germany.
}

\date{Accepted XXX. Received YYY; in original form ZZZ}

\pubyear{2015}

\begin{document}
\label{firstpage}
\pagerange{\pageref{firstpage}--\pageref{lastpage}}
\maketitle

\begin{abstract}
We present a flux density study of 89 millisecond pulsars (MSPs) regularly monitored as part of the MeerKAT Pulsar Timing Array (MPTA) using the L-Band receiver with an approximately two week cadence between 2019-2022.
For each pulsar, we have determined the mean flux densities
at each epoch in eight $\sim$97 MHz sub-bands ranging from 944 to 1625 MHz.
From these we have derived their modulation indices, their average and
peak-to-median flux densities in each sub-band, as well as their mean spectral indices across the entire frequency range. 
We find that the vast majority of the MSPs have spectra that are well described by a simple power law, with a mean spectral index of --1.86(6).
Using the temporal variation of the flux densities we measured the structure functions and determined the refractive scintillation timescale for seven. The structure functions provide strong evidence that the intrinsic radio luminosities of MSPs are stable.
As a population, the average modulation index at 20\,cm wavelengths peaks near unity
at dispersion measures 
(DMs) of $\sim$20 pc cm$^{-3}$ and by a DM of 100 pc cm$^{-3}$ are closer to 0.2, due to refractive scintillation. 
We find that timing arrays can improve their observing efficiency
by reacting to scintillation maxima, and that
20\,cm FRB surveys should prioritise highly scintillating mid-latitude 
regions of the Galactic sky where they will find $\sim$30\% more events and bursts at greater distances. 
\end{abstract}




\section{Introduction}
\label{sec:intro}

\par 
Timing millisecond pulsars has
a large number
of astrophysical applications,
including the mass measurements of neutron stars \citep{1999ApJ...512..288T} and the pursuit of nanohertz frequency gravitational wave detection 
using arrays of precisely timed MSPs known as pulsar timing arrays (PTAs) \citep{1979ApJ...234.1100D, 1983ApJ...265L..39H}. 

PTAs \citep[e.g.][]{2023ApJ...951L...9A, 2023arXiv230616230Z, 2023arXiv230616224A} observe the same MSPs over many years, 
providing extended and comprehensive data sets that can be used for a wide range of investigations including polarimetry and flux density monitoring \citep{1999ApJS..123..627S, 1999ApJ...526..957K, 2004MNRAS.352..804O, 2015MNRAS.449.3223D}. 

Radio astronomers often use catalogued flux
densities, pulse shapes and spectral indices to predict the timing precision they can achieve in a given
time with their radio telescope based upon the radiometer equation \citep[e.g.][]{2010IAUS..261..212B}.
The pulsar population's flux densities and pulse shapes can also be used to estimate
the total galactic pulsar population \citep{2013MNRAS.434.1387L} and the yields of future
surveys with instruments such as the Square Kilometre Array (SKA) \citep{2009A&A...493.1161S, 2015aska.confE..40K, 2017PASA...34...70X}.

The majority of radio pulsars have been discovered in large area 
unbiased surveys and it is common practice to publish their mean radio
flux densities along with their rotational, binary and astrometric parameters \citep[e.g.][]{2001MNRAS.328...17M, 2013ApJ...763...80B}. 
Some survey papers use simple calibrations that rely upon the radiometer equation to
estimate the flux density of the newly discovered pulsars \citep{2001MNRAS.328...17M}, whilst others usually
use more reliable calibration procedures involving 
a pulsed calibrator signal combined with observations of a reference radio source
such as Hydra A \citep{2021MNRAS.504..228S}.
Some pulsars that are being routinely observed as part of timing programs
provide flux densities that ultimately appear in the pulsar catalogue.
Version 1.69 of the ATNF radio pulsar catalog \citep{2005AJ....129.1993M} lists 
flux density estimates for
2344 pulsars at a radio frequency of 1400 MHz.
Unfortunately, the mean flux densities of pulsars in the literature often vary.
This is partly because many pulsars exhibit variations in their flux densities between observing epochs due to interstellar scintillation, but possibly also because of the methodologies and calibration procedures used.

Radio pulsar flux densities vary for many reasons and on different
timescales. Indeed, confirmation of the first pulsar discovered (B1919+21) \citep{1968Natur.217..709H}, failed
on the first attempt due to unfavourable interstellar scintillation
that deamplified the pulsar's flux density resulting in a non-detection.

Flux density variations can be classified as due to intrinsic or extrinsic reasons.

Here we define intrinsic causes as those which do not depend upon the nature
of the bulk of the interstellar medium between the observer and the pulsar 
along the line of sight. These include magnetospheric changes that lead
to both pulse-to-pulse variations, and nulling.
Although the mean pulse profiles of pulsars are usually stable, their
individual pulses vary in amplitude and shape. These pulse-to-pulse variations  
are sometimes known as pulse `jitter' \citep{1985ApJS...59..343C,2022MNRAS.510.5908M}.  
Some pulsars often switch between two or more distinct mean pulse profiles or radio emission states known as mode changing \citep{1970Natur.228.1297B} and in another observed phenomena the radio pulses can cease completely for a few to many tens of rotation periods, a phenomena known as pulse nulling \citep{1970Natur.228...42B}. 
A special class of pulsars known as intermittent pulsars behave as regular pulsars for certain intervals of time and then turn off completely for what can be months \citep{2006Sci...312..549K}.
In this paper we will
only consider the pulsar flux density that is the average of many 100s to 1000s of rotations
and thus ignores flux density variations due to pulse jitter. In rare circumstances, precession of a pulsar's spin axis due to relativistic spin-orbit coupling can change our line of sight through the emission cone and change the flux density of the pulsar. 
The best known example of this was exhibited by the B pulsar in the double pulsar which completely faded from view and is now invisible \citep{2021PhRvX..11d1050K}.
The final form of intrinsic variation arises from eclipses of the pulsar by
a companion. This can lead to either severe attenuation or complete non-detection of the pulsar
such as in the case of the eclipsing pulsar PSR B1259--63 that orbits a Be star \citep{1992ApJ...387L..37J}. 

\par Radio pulsars also experience changes in their mean flux densities because of the propagation of
their radiation through the ionised interstellar medium (IISM) which we consider to
be extrinsic in nature.
The IISM causes the radio emission to experience refractive and diffractive
scintillation that change the phase and path length of the radiation on its
way to the Earth. 
The interference of the radio waves can result in strong flux density variations in both frequency and time, on a timescale between a 
few seconds to several hours, known as diffractive scintillation. Several studies have been performed to study diffractive scintillation effects, which have been often interpreted successfully 
 using the `thin screen scattering' model \citep{1969Natur.221..158R, 1986ApJ...310..737C, 1995ApJ...443..209A}. Using the derived diffractive scintillation parameters of bandwidth and timescale as a function of
 both day of year and orbital phase, a pulsar's mean transverse velocity and orientation of the orbit along our line of sight can be estimated \citep{1986ApJ...311..183C, 2020ApJ...904..104R, 2023MNRAS.519.5086A}. 
 
 Refractive scintillation is the process of refractive focussing and de-focussing of the pulsar emission due to larger structure electron density fluctuations along our line of sight in the IISM. This causes the flux densities to vary slowly over a long period of time ranging from days to months. The longer timescales involved make studies of refractive scintillation more challenging and involve multi-year campaigns.
  The long-term variations of pulsar flux densities were first studied by \cite{1970Natur.225..712C} with the theoretical framework being developed by \cite{1968Natur.218..920S}, \cite{1984A&A...134..390R} and \cite{1986ApJ...310..737C}.
  In a 43 consecutive day mean flux density campaign on 25 pulsars with the NRAO 91 m 
  telescope, \cite{1990ApJ...352..207S} demonstrated for the first time that potentially all of the
  non-nulling ``normal'' slow pulsars with high dispersion measures had remarkably consistent mean flux densities and that their intrinsic luminosities are probably constant. This was corroborated by
  a more comprehensive study consisting of daily monitoring 
  over five years of 21 pulsars at a centre frequency of 610 MHz \citep{2000ApJ...539..300S}. The pulsars with the highest
  DMs ($>200$\,pc cm$^{-3}$) showed remarkably low levels of flux modulation
  of just a few percent.
  More recently
  \cite{2021MNRAS.501.4490K} studied the effects of the IISM by long term monitoring of flux densities of 286 normal pulsars and estimated various scintillation timescales involved. Again they found that
  the majority of pulsars with large DMs had low
  modulation indices, consistent with constant intrinsic radio luminosities.
  Their only exceptions appeared to be subject to obvious nulling or eclipsing
  behaviour.

Millisecond pulsars are steep spectrum objects, and at 20\,cm wavelengths often
have power law spectra with power law indices most often between $-$1 to $-$3. \cite{1998ApJ...501..270K} monitored 23 MSPs with the 100\,m Effelsberg radio telescope and found their mean spectral index was $-1.8$(1) whilst \cite{1999ApJ...526..957K} found the mean spectral index of $-1.76(14) $ for a wider range of frequencies extending up to $ \sim 4.9$ GHz.
\cite{1998ApJ...506..863T} performed a comprehensive study of pulsar flux densities to explore the spectra of 19 MSPs with the Parkes 64\,m radio telescope using three different receivers.
They found that the MSPs in their study had spectral indices that ranged between $-1.1$ for PSR J0437--4715 and $-2.9$ for PSR J1804--2718 with a mean of $-1.9(1)$. 
Recently \cite{2018MNRAS.473.4436J} explored the
flux densities and spectral properties of 441 pulsars. They found that 70\% of
all the pulsars in their study could have their spectra defined by a simple power law, but others
were better modelled with a spectral break, whilst others needed a curved
spectrum. In a study of the mean flux densities of 24 MSPs in three
frequency bands \cite{2015MNRAS.449.3223D} showed that
most MSPs were also well described by a single power law spectra, with a few exceptions such as PSR J1600--3053 (which shows a flattening of its spectrum below 1 GHz).
Most recently \cite{2022PASA...39...27S} measured the spectral
indices of 189 MSPs with the MeerKAT radio telescope 
and found a mean spectral index of $-1.92$(6), very close to other mean values in the literature.

Stable flux density measurements of the high-DM slow pulsars suggests that pulsars have intrinsically stable radio luminosities over long timescales and the observed variations in the flux densities can be attributed to fluctuations in the turbulent plasma in the IISM. This may not be very surprising. Most pulsars are millions
of years old and have only been observed for a few tens of years. If they have stable magnetic
fields and orientations, they might be expected to be intrinsically stable radio emitters.
One way to characterise the changes in flux is to define a refractive scintillation timescale that
reflects the timescale over which pulsar fluxes are correlated.
 \cite{2000ApJ...539..300S} and more recently \cite{2023MNRAS.tmp..221W} studied the
 flux density variations of 21 and 151 pulsars respectively.
 This enabled them to estimate the refractive scintillation timescales of 21 and 15 pulsars respectively, and \cite{2023MNRAS.tmp..221W} measured the modulation indices of 95 pulsars and
 39 of their pulsars had a pulse period less than 20 ms.
 Although their study only included 14 MSPs with DMs greater than 100 pc cm$^{-3}$
 they possessed a mean modulation index (standard deviation / mean flux density) of 0.23. This
 suggests that any intrinsic variations are less than this and therefore, like many of the slow pulsars,
 MSPs might also have stable intrinsic radio luminosities.

The new 64-antenna MeerKAT telescope located in the Karoo, South Africa provides an opportunity to revisit and monitor the flux densities of MSPs in the southern sky \citep{2020PASA...37...28B}. 
The MeerKAT Pulsar Timing Array (MPTA) is one of four major project themes in the MeerTime project
\citep{2022PASA...39...27S, 2023MNRAS.519.3976M}.
The principal goal of the MPTA is to detect nanohertz frequency gravitational waves (GWs).

Long term monitoring of MSP flux densities can contribute to optimising the observation strategy for the faster detection of GWs. If pulsars experience long-term refractive scintillation 
that leads to predictable fluxes over many months
we may choose to dwell longer on them when they are bright, and spend less time observing them 
when they are faint. 
Decades of MSP flux density monitoring by the PTAs will also assist in gaining a better understanding of the IISM if their radio luminosities are intrinsically stable.

\par In this work, we measured the flux density variability of the 89 MSPs that are part of the MPTA 
program across nearly an octave of bandwidth with a centre frequency
near 1284 MHz over many epochs spanning 3.3 years.
In Section~\ref{sec:Psr_sample}, a brief summary of the pulsar sample 
and observations is provided before we discuss 
the data reduction and flux density calibration process in
Section~\ref{subsec:flux_cal}. In Section~\ref{sec:res}, we 
present our results, including the mean flux densities, spectral
indices, modulation indices, and temporal flux density structure functions.
In Section \ref{sec:discussion}, we look at some of the aggregate properties of the pulsar sample
and their implications for timing arrays and FRBs discovered at high galactic latitudes \citep{2013Sci...341...53T}.
Finally, we draw our conclusions in Section~\ref{sec:conclusion}.


\begin{table*}
\centering
\caption{Measured flux densities of the 89 MSPs in our sample. $N_{\rm obs}$ is the number of observations for each pulsar and $S_{944}$, $m_{944} $, $ R_{944}$ are the mean flux density, modulation index and ratio of the maximum to median value of the flux density at a central frequency of 944~MHz respectively. Similarly these quantities are given for the 1429 MHz and 1625 MHz sub-bands of the observing frequency band. Spectral index is calculated here using a simple power law model fit to the mean flux density values across the 8 frequency sub-bands. See the data availability section for the results for all 8 of the sub-bands.}
\label{tab:mod_ind}

\begin{tabular}{rrrrrrrrrrrrr}
\hline
 
 \multicolumn{1}{c}{NAME}      &  \multicolumn{1}{c}{$ N_{\rm obs}$} &  \multicolumn{1}{c}{DM}       &  \multicolumn{1}{c}{$ S_{944}$}       &  \multicolumn{1}{c}{$ m_{944}$}      &  \multicolumn{1}{c}{$R_{944}$} &  \multicolumn{1}{c}{$S_{1429}$}       &  \multicolumn{1}{c}{$ m_{1429} $}      &  \multicolumn{1}{c}{$R_{1429}$} &  \multicolumn{1}{c}{$ S_{1625}$}       &  \multicolumn{1}{c}{$ m_{1625} $}      &  \multicolumn{1}{c}{$R_{1625}$} &  \multicolumn{1}{c}{Spectral}   \\
        &    &(pc cm$^{-3}$) & \multicolumn{1}{c}{(mJy)} &   & & \multicolumn{1}{c}{(mJy)} &     &  & \multicolumn{1}{c}{(mJy)} &    &  &  \multicolumn{1}{c}{Index} \\
\hline  

J0030+0451 & 67   & 4.3   & 2.7(2)    & 0.7(1)    & 5.4     & 1.12(5)    & 0.39(5)    & 2.6     & 0.83(3)    & 0.34(4)    & 2.3     & --2.1(1)           \\
J0101--6422 & 39   & 11.9  & 0.50(10)   & 1.2(4)    & 10.6    & 0.24(2)    & 0.5(1)    & 2.6     & 0.17(1)    & 0.5(1)    & 2.9     & --1.7(2)           \\
J0125--2327 & 76   & 9.6   & 4.6(4)    & 0.8(1)    & 4.6     & 3.2(2)    & 0.7(1)    & 3.5     & 2.5(2)    & 0.61(9)    & 3.4     & --1.2(2)           \\
J0437--4715 & 109  & 2.6   & 265(11)   & 0.45(5)    & 2.6     & 137(4)    & 0.30(3)    & 2.0     & 104(3)    & 0.27(2)    & 1.9     & --1.76(7)           \\
J0610--2100 & 73   & 60.7  & 1.05(4)    & 0.35(4)    & 2.0     & 0.65(4)    & 0.59(9)    & 4.4     & 0.57(4)    & 0.7(1)    & 5.3     & --1.1(1)           \\
J0613--0200 & 71   & 38.8  & 4.5(1)    & 0.28(3)    & 1.8     & 2.01(9)    & 0.38(5)    & 3.0     & 1.51(7)    & 0.41(5)    & 2.9     & --1.97(8)           \\
J0614--3329 & 88   & 37.1  & 1.48(7)    & 0.43(5)    & 3.2     & 0.66(6)    & 0.8(1)    & 7.2     & 0.51(4)    & 0.8(1)    & 5.4     & --1.9(1)           \\
J0636--3044 & 56   & 15.5  & 2.3(5)    & 1.6(5)    & 19.9    & 1.5(2)    & 1.2(3)    & 9.0     & 1.3(2)    & 1.0(2)    & 7.2     & --0.9(3)           \\
J0711--6830 & 76   & 18.4  & 6.0(7)    & 1.0(2)    & 7.5     & 3.2(6)    & 1.6(4)    & 20.7    & 2.1(3)    & 1.3(3)    & 11.1    & --1.6(3)           \\
J0900--3144 & 76   & 75.7  & 5.84(7)    & 0.098(9)    & 1.3     & 3.67(5)    & 0.11(1)    & 1.2     & 2.94(4)    & 0.12(1)    & 1.3     & --1.25(2)           \\
J0931--1902 & 59   & 41.5  & 1.5(1)    & 0.7(1)    & 4.6     & 0.49(6)    & 0.9(2)    & 7.5     & 0.40(6)    & 1.1(3)    & 12.2    & --2.4(2)           \\
J0955--6150 & 103  & 160.9 & 2.25(5)    & 0.24(2)    & 1.8     & 0.63(2)    & 0.32(3)    & 2.1     & 0.40(2)    & 0.42(5)    & 2.2     & --3.13(6)           \\
J1012--4235 & 82   & 71.7  & 0.42(1)    & 0.28(3)    & 1.8     & 0.27(1)    & 0.34(4)    & 2.3     & 0.23(1)    & 0.42(5)    & 2.8     & --1.12(7)           \\
J1017--7156 & 89   & 94.2  & 2.28(9)    & 0.37(4)    & 2.3     & 1.00(5)    & 0.43(5)    & 2.7     & 0.73(4)    & 0.46(6)    & 2.8     & --2.06(8)           \\
J1022+1001 & 71   & 10.3  & 6.3(8)    & 1.1(2)    & 8.3     & 4.2(4)    & 0.9(2)    & 5.6     & 3.2(3)    & 0.8(1)    & 4.9     & --1.6(2)           \\
J1024--0719 & 73   & 6.5   & 3.2(6)    & 1.5(4)    & 23.2    & 1.5(1)    & 0.7(1)    & 4.7     & 1.14(9)    & 0.7(1)    & 3.9     & --1.9(2)           \\
J1036--8317 & 82   & 27.1  & 0.69(4)    & 0.49(7)    & 2.8     & 0.41(4)    & 0.8(1)    & 5.5     & 0.32(3)    & 0.8(1)    & 4.0     & --1.3(1)           \\
J1045--4509 & 77   & 58.1  & 5.8(2)    & 0.26(3)    & 1.7     & 2.35(9)    & 0.33(4)    & 2.1     & 1.74(7)    & 0.37(4)    & 2.2     & --2.19(7)           \\
J1101--6424 & 86   & 207.4 & 0.59(1)    & 0.17(2)    & 1.5     & 0.309(6)    & 0.18(2)    & 1.5     & 0.243(6)    & 0.24(2)    & 1.7     & --1.63(4)           \\
J1103--5403 & 78   & 103.9 & 1.02(7)    & 0.57(8)    & 3.0     & 0.39(2)    & 0.54(8)    & 2.8     & 0.30(2)    & 0.61(9)    & 3.8     & --2.3(1)           \\
J1125--5825 & 77   & 124.8 & 1.59(4)    & 0.24(3)    & 1.8     & 0.97(3)    & 0.23(2)    & 1.8     & 0.80(2)    & 0.22(2)    & 1.6     & --1.27(5)           \\
J1125--6014 & 88   & 52.9  & 2.23(5)    & 0.22(2)    & 1.8     & 1.35(5)    & 0.37(4)    & 2.5     & 1.05(5)    & 0.42(5)    & 2.7     & --1.30(6)           \\
J1216--6410 & 77   & 47.4  & 2.62(5)    & 0.18(2)    & 1.5     & 1.21(4)    & 0.26(3)    & 1.9     & 0.86(3)    & 0.26(3)    & 1.7     & --1.99(5)           \\
J1231--1411 & 49   & 8.1   & 0.7(1)    & 1.0(2)    & 8.5     & 0.39(5)    & 0.9(2)    & 8.1     & 0.28(4)    & 0.9(2)    & 8.3     & --1.7(3)           \\
J1327--0755 & 59   & 27.9  & 0.51(7)    & 1.0(2)    & 8.1     & 0.17(2)    & 1.1(3)    & 8.4     & 0.16(3)    & 1.7(5)    & 18.3    & --2.5(3)           \\
J1421--4409 & 74   & 54.6  & 2.4(1)    & 0.40(5)    & 2.3     & 1.33(7)    & 0.43(6)    & 2.7     & 0.95(5)    & 0.48(7)    & 3.1     & --1.64(10)          \\
J1431--5740 & 75   & 131.4 & 0.58(2)    & 0.25(3)    & 2.0     & 0.38(1)    & 0.26(3)    & 2.4     & 0.294(8)    & 0.24(3)    & 2.0     & --1.28(6)           \\
J1435--6100 & 84   & 113.8 & 0.524(8)    & 0.13(1)    & 1.3     & 0.311(7)    & 0.20(2)    & 1.6     & 0.241(6)    & 0.23(2)    & 1.7     & --1.35(4)           \\
J1446--4701 & 74   & 55.8  & 0.91(6)    & 0.52(7)    & 3.0     & 0.37(3)    & 0.7(1)    & 3.8     & 0.26(3)    & 0.9(2)    & 6.3     & --2.1(2)           \\
J1455--3330 & 73   & 13.6  & 2.2(3)    & 1.3(3)    & 11.3    & 0.72(9)    & 1.1(2)    & 9.5     & 0.50(5)    & 0.9(2)    & 5.1     & --2.8(3)           \\
J1513--2550 & 37   & 46.9  & 1.61(7)    & 0.28(5)    & 1.4     & 0.36(1)    & 0.25(4)    & 1.7     & 0.22(1)    & 0.33(6)    & 1.6     & --3.65(10)          \\
J1514--4946 & 44   & 31.0  & 0.27(3)    & 0.7(1)    & 4.2     & 0.19(2)    & 0.8(2)    & 4.5     & 0.15(2)    & 1.1(3)    & 8.8     & --1.0(3)           \\
J1525--5545 & 79   & 127.0 & 0.72(1)    & 0.13(1)    & 1.5     & 0.426(6)    & 0.13(1)    & 1.4     & 0.337(5)    & 0.13(1)    & 1.5     & --1.40(3)           \\
J1543--5149 & 75   & 51.0  & 3.2(1)    & 0.36(4)    & 2.4     & 0.87(6)    & 0.57(9)    & 4.4     & 0.60(4)    & 0.63(10)   & 4.0     & --3.1(1)           \\
J1545--4550 & 84   & 68.4  & 1.64(6)    & 0.35(4)    & 2.1     & 1.03(4)    & 0.38(4)    & 2.6     & 0.92(4)    & 0.44(5)    & 2.5     & --1.08(8)           \\
J1547--5709 & 73   & 95.7  & 0.68(3)    & 0.36(4)    & 2.0     & 0.34(1)    & 0.32(4)    & 1.9     & 0.26(1)    & 0.36(5)    & 1.9     & --1.76(8)           \\
J1600--3053 & 73   & 52.3  & 3.1(1)    & 0.28(3)    & 1.7     & 2.30(8)    & 0.29(3)    & 1.9     & 1.94(7)    & 0.32(4)    & 2.0     & --0.88(7)           \\
J1603--7202 & 77   & 38.0  & 7.1(5)    & 0.64(10)   & 5.5     & 2.6(2)    & 0.8(1)    & 4.7     & 1.9(2)    & 0.9(2)    & 6.3     & --2.4(2)           \\
J1614--2230 & 76   & 34.5  & 2.6(1)    & 0.36(5)    & 2.2     & 1.12(6)    & 0.46(6)    & 2.3     & 0.92(6)    & 0.56(8)    & 3.7     & --1.9(1)           \\
J1628--3205 & 13   & 42.1  & 1.4(1)    & 0.4(1)    & 1.9     & 0.48(7)    & 0.6(2)    & 2.8     & 0.39(7)    & 0.7(3)    & 3.2     & --2.5(2)           \\
J1629--6902 & 79   & 29.5  & 2.28(10)   & 0.38(5)    & 2.3     & 0.96(7)    & 0.63(10)   & 4.6     & 0.70(6)    & 0.7(1)    & 5.5     & --2.1(1)           \\
J1643--1224 & 76   & 62.4  & 9.0(1)    & 0.11(1)    & 1.3     & 3.94(6)    & 0.13(1)    & 1.3     & 2.99(6)    & 0.18(2)    & 1.4     & --2.00(3)           \\
J1652--4838 & 75   & 188.2 & 1.46(3)    & 0.16(2)    & 1.3     & 0.99(2)    & 0.17(2)    & 1.6     & 0.84(2)    & 0.21(2)    & 1.6     & --1.01(4)           \\
J1653--2054 & 61   & 56.5  & 1.35(6)    & 0.38(5)    & 2.4     & 0.55(3)    & 0.38(5)    & 2.2     & 0.41(2)    & 0.40(6)    & 2.2     & --2.17(10)          \\
J1658--5324 & 50   & 30.8  & 1.2(1)    & 0.6(1)    & 3.8     & 0.59(9)    & 1.0(3)    & 7.9     & 0.38(5)    & 0.9(2)    & 6.0     & --2.1(2)           \\
J1705--1903 & 60   & 57.5  & 1.15(6)    & 0.39(6)    & 1.7     & 0.69(3)    & 0.36(5)    & 1.9     & 0.54(2)    & 0.35(5)    & 1.9     & --1.40(10)          \\
J1708--3506 & 61   & 146.8 & 3.46(5)    & 0.11(1)    & 1.2     & 1.46(2)    & 0.10(1)    & 1.2     & 1.04(2)    & 0.18(2)    & 1.2     & --2.16(3)           \\
J1713+0747 & 74   & 16.0  & 9.7(8)    & 0.7(1)    & 4.6     & 6.9(9)    & 1.1(2)    & 10.5    & 5.7(7)    & 1.0(2)    & 10.5    & --1.0(2)           \\
J1719--1438 & 76   & 36.8  & 1.09(6)    & 0.48(7)    & 3.2     & 0.36(3)    & 0.7(1)    & 4.8     & 0.29(4)    & 1.3(3)    & 14.4    & --2.6(2)           \\
J1721--2457 & 48   & 48.2  & 2.11(8)    & 0.25(4)    & 1.5     & 1.09(7)    & 0.43(7)    & 2.2     & 0.81(5)    & 0.40(6)    & 2.2     & --1.67(10)          \\
J1730--2304 & 73   & 9.6   & 8.0(5)    & 0.48(7)    & 3.1     & 3.8(4)    & 0.9(2)    & 8.9     & 2.7(4)    & 1.2(3)    & 13.4    & --1.9(2)           \\
J1731--1847 & 14   & 106.5 & 1.02(4)    & 0.14(3)    & 1.2     & 0.38(2)    & 0.23(6)    & 1.2     & 0.30(2)    & 0.26(7)    & 1.3     & --2.3(1)           \\
J1732--5049 & 87   & 56.8  & 4.2(3)    & 0.61(9)    & 4.4     & 2.0(2)    & 0.9(1)    & 6.0     & 1.4(1)    & 0.9(1)    & 7.1     & --1.9(2)           \\
J1737--0811 & 79   & 55.3  & 2.46(7)    & 0.25(3)    & 1.5     & 1.10(3)    & 0.26(3)    & 1.7     & 0.86(3)    & 0.35(4)    & 2.1     & --1.99(6)           \\
J1744--1134 & 75   & 3.1   & 6.5(8)    & 1.1(2)    & 9.5     & 3.1(4)    & 1.0(2)    & 6.0     & 2.5(3)    & 1.0(2)    & 6.4     & --1.5(2)           \\
J1747--4036 & 75   & 152.9 & 3.89(9)    & 0.19(2)    & 1.5     & 1.42(2)    & 0.15(1)    & 1.4     & 0.98(3)    & 0.25(3)    & 1.7     & --2.50(4)           \\
J1751--2857 & 74   & 42.8  & 0.78(2)    & 0.23(3)    & 1.6     & 0.47(1)    & 0.27(3)    & 1.8     & 0.37(1)    & 0.31(4)    & 2.2     & --1.37(6)           \\
J1756--2251 & 95   & 121.2 & 1.70(2)    & 0.13(1)    & 1.2     & 1.03(2)    & 0.19(2)    & 1.4     & 0.83(2)    & 0.21(2)    & 1.4     & --1.30(3)           \\
J1757--5322 & 93   & 30.8  & 3.2(1)    & 0.33(4)    & 2.1     & 1.70(9)    & 0.49(6)    & 2.7     & 1.4(1)    & 0.8(1)    & 8.2     & --1.64(10)          \\
J1801--1417 & 73   & 57.3  & 3.2(1)    & 0.34(4)    & 2.0     & 1.62(7)    & 0.38(5)    & 2.6     & 1.26(6)    & 0.41(5)    & 2.6     & --1.70(9)           \\

\hline
\end{tabular}
\end{table*}

\begin{table*}
\centering
\begin{tabular}{rrrrrrrrrrrrr}
\hline

 \multicolumn{1}{c}{NAME}      &  \multicolumn{1}{c}{$ N_{\rm obs}$} &  \multicolumn{1}{c}{DM}       &  \multicolumn{1}{c}{$ S_{944}$}       &  \multicolumn{1}{c}{$ m_{944}$}      &  \multicolumn{1}{c}{$R_{944}$} &  \multicolumn{1}{c}{$S_{1429}$}       &  \multicolumn{1}{c}{$ m_{1429} $}      &  \multicolumn{1}{c}{$R_{1429}$} &  \multicolumn{1}{c}{$ S_{1625}$}       &  \multicolumn{1}{c}{$ m_{1625} $}      &  \multicolumn{1}{c}{$R_{1625}$} &  \multicolumn{1}{c}{Spectral}   \\
        &    &(pc cm$^{-3}$) & \multicolumn{1}{c}{(mJy)} &   & & \multicolumn{1}{c}{(mJy)} &     &  & \multicolumn{1}{c}{(mJy)} &    &  &  \multicolumn{1}{c}{Index} \\
\hline 
J1802--2124 & 72   & 149.6 & 2.06(4)    & 0.16(2)    & 1.4     & 0.75(2)    & 0.18(2)    & 1.5     & 0.53(1)    & 0.20(2)    & 1.5     & --2.47(4)      \\
J1804--2717 & 36   & 24.7  & 3.7(3)    & 0.5(1)    & 2.4     & 1.8(4)    & 1.2(4)    & 11.0    & 1.3(2)    & 1.1(3)    & 7.8     & --1.8(3)           \\
J1804--2858 & 45   & 232.5 & 1.38(4)    & 0.18(2)    & 1.4     & 0.94(2)    & 0.12(1)    & 1.2     & 0.66(2)    & 0.18(2)    & 1.4     & --1.46(5)           \\
J1811--2405 & 119  & 60.6  & 3.11(6)    & 0.20(2)    & 1.5     & 1.44(3)    & 0.22(2)    & 1.7     & 1.10(3)    & 0.26(2)    & 1.8     & --1.90(4)           \\
J1825--0319 & 73   & 119.6 & 0.364(9)    & 0.21(2)    & 1.4     & 0.206(4)    & 0.18(2)    & 1.3     & 0.162(5)    & 0.26(3)    & 1.6     & --1.51(5)           \\
J1832--0836 & 54   & 28.2  & 1.8(1)    & 0.49(8)    & 3.5     & 1.0(1)    & 0.8(2)    & 6.7     & 0.69(5)    & 0.6(1)    & 3.2     & --1.7(2)           \\
J1843--1113 & 73   & 60.0  & 1.17(5)    & 0.37(5)    & 2.0     & 0.55(3)    & 0.47(7)    & 2.5     & 0.47(4)    & 0.7(1)    & 5.6     & --1.7(1)           \\
J1843--1448 & 38   & 114.5 & 0.99(1)    & 0.08(1)    & 1.1     & 0.515(9)    & 0.11(1)    & 1.2     & 0.406(8)    & 0.13(2)    & 1.3     & --1.70(3)           \\
J1902--5105 & 78   & 36.3  & 3.29(6)    & 0.15(2)    & 1.4     & 0.96(2)    & 0.19(2)    & 1.7     & 0.64(2)    & 0.23(2)    & 1.9     & --3.02(4)           \\
J1903--7051 & 90   & 19.7  & 1.9(2)    & 0.8(1)    & 5.6     & 0.9(1)    & 1.1(2)    & 8.8     & 0.75(9)    & 1.1(2)    & 10.4    & --1.8(2)           \\
J1909--3744 & 192  & 10.4  & 3.6(2)    & 0.86(10)   & 7.2     & 1.9(2)    & 1.3(2)    & 18.0    & 1.6(2)    & 2.0(4)    & 46.7    & --1.5(2)           \\
J1911--1114 & 39   & 31.0  & 2.6(2)    & 0.6(1)    & 3.0     & 0.92(10)   & 0.7(2)    & 3.7     & 0.68(9)    & 0.9(2)    & 5.4     & --2.5(2)           \\
J1918--0642 & 76   & 26.6  & 3.5(1)    & 0.27(3)    & 2.2     & 1.79(8)    & 0.38(5)    & 2.1     & 1.22(9)    & 0.61(9)    & 4.2     & --1.84(9)           \\
J1933--6211 & 95   & 11.5  & 2.1(3)    & 1.2(2)    & 12.9    & 1.0(1)    & 1.4(3)    & 18.1    & 0.8(1)    & 1.3(3)    & 12.6    & --1.8(3)           \\
J1946--5403 & 75   & 23.7  & 0.58(9)    & 1.3(3)    & 12.0    & 0.30(5)    & 1.6(4)    & 20.4    & 0.27(4)    & 1.4(3)    & 13.5    & --1.3(3)           \\
J2010--1323 & 77   & 22.2  & 1.32(4)    & 0.28(3)    & 2.0     & 0.67(4)    & 0.46(6)    & 3.2     & 0.56(4)    & 0.58(8)    & 3.3     & --1.59(9)           \\
J2039--3616 & 74   & 24.0  & 1.1(1)    & 1.1(2)    & 11.8    & 0.50(7)    & 1.2(3)    & 12.5    & 0.40(7)    & 1.5(4)    & 14.7    & --1.7(3)           \\
J2124--3358 & 77   & 4.6   & 11(1)    & 1.0(2)    & 7.1     & 4.9(3)    & 0.57(8)    & 3.5     & 3.5(2)    & 0.52(7)    & 2.7     & --2.2(2)           \\
J2129--5721 & 81   & 31.8  & 3.8(5)    & 1.1(2)    & 11.3    & 0.9(1)    & 1.2(3)    & 14.5    & 0.58(10)   & 1.5(4)    & 21.9    & --3.4(3)           \\
J2145--0750 & 76   & 9.0   & 17(2)    & 1.2(2)    & 10.7    & 5.7(8)    & 1.2(3)    & 12.7    & 4.6(5)    & 1.0(2)    & 7.5     & --2.2(3)           \\
J2150--0326 & 70   & 20.7  & 0.99(8)    & 0.7(1)    & 4.8     & 0.37(5)    & 1.1(2)    & 8.4     & 0.30(4)    & 1.1(2)    & 10.7    & --2.3(2)           \\
J2222--0137 & 74   & 3.3   & 2.1(3)    & 1.0(2)    & 9.5     & 1.10(10)   & 0.7(1)    & 4.5     & 0.95(8)    & 0.7(1)    & 4.2     & --1.6(2)           \\
J2229+2643 & 66   & 22.7  & 1.5(2)    & 1.3(3)    & 14.8    & 1.0(3)    & 2.0(7)    & 32.1    & 0.6(1)    & 1.5(4)    & 14.9    & --1.3(4)           \\
J2234+0944 & 69   & 17.8  & 2.7(5)    & 1.6(4)    & 20.0    & 1.6(2)    & 1.1(3)    & 11.6    & 1.1(1)    & 1.1(2)    & 10.7    & --1.8(3)           \\
J2236--5527 & 49   & 20.1  & 0.63(8)    & 0.9(2)    & 5.9     & 0.25(4)    & 1.2(3)    & 8.1     & 0.22(4)    & 1.2(3)    & 7.5     & --2.0(3)           \\
J2241--5236 & 84   & 11.4  & 4.9(6)    & 1.1(2)    & 9.4     & 1.7(1)    & 0.61(9)    & 4.3     & 1.16(7)    & 0.54(7)    & 3.8     & --2.8(2)           \\
J2317+1439 & 68   & 21.9  & 1.6(2)    & 0.8(1)    & 4.6     & 0.48(8)    & 1.3(3)    & 14.5    & 0.40(9)    & 2.0(6)    & 35.1    & --2.8(3)           \\
J2322+2057 & 55   & 13.4  & 0.70(9)    & 0.9(2)    & 5.9     & 0.34(5)    & 1.1(3)    & 9.5     & 0.26(5)    & 1.3(4)    & 14.5    & --1.9(3)           \\
J2322--2650 & 71   & 6.1   & 0.38(4)    & 1.0(2)    & 7.0     & 0.24(1)    & 0.50(7)    & 2.7     & 0.19(1)    & 0.50(7)    & 2.8     & --1.3(2)  \\        

\hline
\end{tabular}
\end{table*}



\section{Observations and Data reduction} \label{sec:Observations and Data reduction}


\subsection{Pulsar sample} \label{sec:Psr_sample}
The MeerKAT radio telescope regularly observes 89 MSPs under the MeerTime MPTA project
\citep{2023MNRAS.519.3976M}. These pulsars were distilled from a longer list of 189 timing array pulsar candidates \citep{2022PASA...39...27S} and chosen for their precision timing potential.
They were observed irregularly at first while the source list was being optimised, and then with an approximately 2-week cadence. Almost all of the pulsars have shown great timing precision potential with mostly sub-microsecond residual timing errors many using only 256\,s long observations \citep{2023MNRAS.519.3976M}. However, one pulsar, PSR J1756--2251, has now been removed from the MPTA project as its
timing precision is not sufficient for retention in the PTA.  
This pulsar was, nevertheless, retained in our flux density study. 

Our data reduction procedure is identical to that described in \cite{2022PASA...39...27S} and a brief summary is presented here. The observations were carried out using the L-band ($856$-$1712$ MHz) receiver of MeerKAT with 1024 frequency channels. The uppermost and lowermost 48 channels of the band were removed to avoid data affected by bandpass roll-off, leaving 775.75\,MHz of bandwidth centred at 1284 MHz. All of the observations then have 928 frequency channels which were integrated into 8 sub-bands for the flux density calculations
after radio frequency interference (RFI) mitigation algorithms were applied. 
The observations typically range from 4-30\,min except for some very long observations ($\sim$ several hours) that were obtained as part of the RelBin project \citep{2021MNRAS.504.2094K} that also observes a subset of the pulsars to study the relativistic effects in binary pulsars. The aim of our experiment was
to explore the flux densities of well-separated
observations (i.e. separated by a day to several days), so we only considered the flux densities
for the first 256\,s of each epoch's observation.
This avoids biases in the flux density distributions 
that can occur if the observations are broken into many
(non-time independent) 4\,min integrations, and averaging
that might occur in long observations.
It would be possible in the future to investigate the
sub-day temporal scintillation properties of the flux densities using the subset of pulsars that have very long observations (i.e. hours) but this is the topic of other studies aiming to
investigate the orbital-phase and day-of-year dependent scintillation properties of some of the relativistic binary pulsars.

Less than $0.5 \%$  of the observations failed due to calibration or
configuration issues and these were removed prior to our analysis.
All the observations use standard templates to help 
identify RFI-affected observations and were cleaned
using the MeerGuard pipeline \citep{2022PASA...39...27S} before integration of the first 256\,s
into our final sample which consists of 8 flux density values per observation. Our data spanned from February 2019 until June 2022. 
On average, the data comprised of $\sim$70 observations per pulsar with a maximum of 192 observations for PSR J1909$-$3744, 
which is often used as a timing calibrator for various timing programs at MeerKAT.  


\subsection{Flux density Calibration}
\label{subsec:flux_cal}
The pulsed cal system at the MeerKAT telescope has non-linearities that make it unsuitable for precise 
polarimetric and flux density calibration, so we instead used the radiometer
equation to estimate the pulsar mean flux densities.
The rms noise $N$ in a radio telescope of gain $G$ over a bandwidth $B$
with $N_{\rm p}$ orthogonal polarisations in an integration time $t$ is given by the radiometer equation
\begin{equation}
\label{eq:radiometer}
    N = {{T_{\rm rec} + T_{\rm sky}} \over{G \sqrt{f BN_{\rm p} t}}}
\end{equation}

\noindent where $f$ is the effective fraction of the bandwidth due to the shape of the polyphase filters employed in the digital signal processing, and $T_{\rm rec}$ and $T_{\rm sky}$ are the temperatures of the receiver, and sky respectively. 
For high precision radiometry, there are additional terms in Eqn \ref{eq:radiometer}, such as effects due to spillover. These are typically small and we can safely ignore them in this work. As the emphasis of this paper is on the relative flux densities and how they vary these assumptions are probably reasonable.
In a folded pulsar observation of $N_{\rm bin}$ phase bins, if there is negligible radio frequency interference, then the pulsar's mean flux density $S$ is related to signal-to-noise ratio $SNR$ 
in the $N_{\rm on}$ on-pulse bins by

\begin{equation}
    S =  SNR {{T_{\rm rec} + T_{\rm sky}}\over{G \sqrt{f BN_{\rm p}t}}} 
    \sqrt{{{N_{\rm on}}\over{N_{\rm bin}-N_{\rm on}}}}.
    \label{eq:flux}
\end{equation}

\noindent The derivation of this equation is given in appendix A. Here $\rm SNR$ should not be confused with the optimal signal-to-noise ratio returned by programs such as {\sc pdmp} from the PSRCHIVE software suite. {\sc pdmp } uses a series of trial pulse widths to find the highest $\rm SNR$ but this neglects the flux in the rest of the on-pulse bins and leads to inaccurate flux densities.
The mean flux density can be computed by integrating
the counts of the on-pulse bins and subtracting off the 
best-fit baseline, 
then determining the average flux density over the entire pulse period. A critical assumption in our analysis was that we
could effectively remove the RFI to validate Eqn \ref{eq:flux}. If so, the MSP flux densities at
large DMs should exhibit low levels of variation (i.e. small modulation indices) if they have radio luminosities
that are intrinsically stable 
from epoch to epoch (which as we will see later, appears
to be the case). Indeed the temporal structure functions of
our pulsar flux densities give us great confidence that our
approach to calibration is valid.

Flux densities were computed using Eqn \ref{eq:flux}. 
We first computed the signal-to-noise ratio $\rm SNR$ of
the counts above our baseline. We then took the sky temperatures from the 408 MHz
all-sky catalogue \citep{1982A&AS...47....1H} at the position of the pulsar
scaled as the frequency $\nu^{-2.6}$ \citep{1987MNRAS.225..307L} to get the sky background temperature. 
To compute the system equivalent flux density $T_{\rm rec}/G$ 
as a function of frequency, we used the polynomial fit 
from SARAO calibrations
presented in \cite{2021MNRAS.505.4468G}. The effective bandwidth 
of each channel and integration time of the observation
was computed after removing
RFI-affected channels and sub-integrations, the loss
of $\rm SNR$ due to the polyphase filterbank ($f=0.91$) \citep{2020PASA...37...28B}
and the number
of on- and off-bins were derived from our standard profiles.  
Each observation was integrated down to 8 frequency sub-bands,
and the flux density determined at each epoch. 
The results are presented in Table \ref{tab:mod_ind}.

\begin{figure*}
\centering
\includegraphics[width=0.8\textwidth]{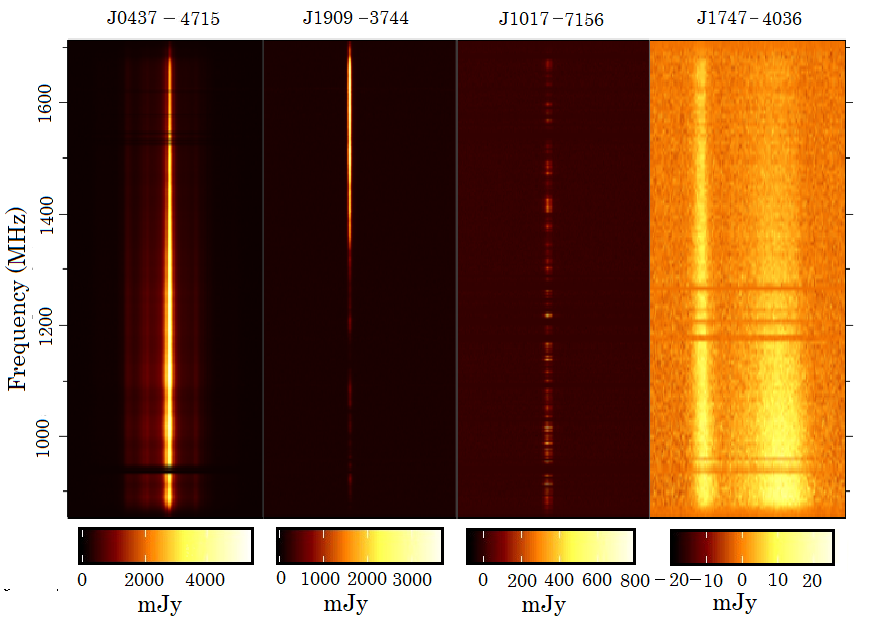}
\caption{The observing frequency versus pulse phase plot of four pulsars at different DMs that experience
very different regimes of interstellar scintillation. The horizontal bands present in some of the plots are due to RFI excision such as that near 950 MHz.
PSR J0437$-$4715 (DM $\sim$ 2.64 pc cm$^{-3}$) is close to us and exhibits broad band scintillation in the 20\,cm band with the flux density never dropping to zero in any part of our observing band. PSR J1909$-$3744 (DM $\sim$ 10.39 pc cm$^{-3}$) often exhibits a sole or a few 
intense scintillation bands whilst PSR J1017$-$7156 (DM $\sim$ 94.22 pc cm$^{-3}$) often exhibits many bright but narrow scintillation bands across the band. 
Finally PSR J1747$-$4036 (DM $\sim$ 152.94 pc cm$^{-3}$) has so many scintillation bands
that they are unresolved at our observing frequencies.}
\label{fig:psr_scint}
\end{figure*}

\begin{figure*}
\centering
    \subcaptionbox{}
    {\includegraphics[width=0.47\textwidth]{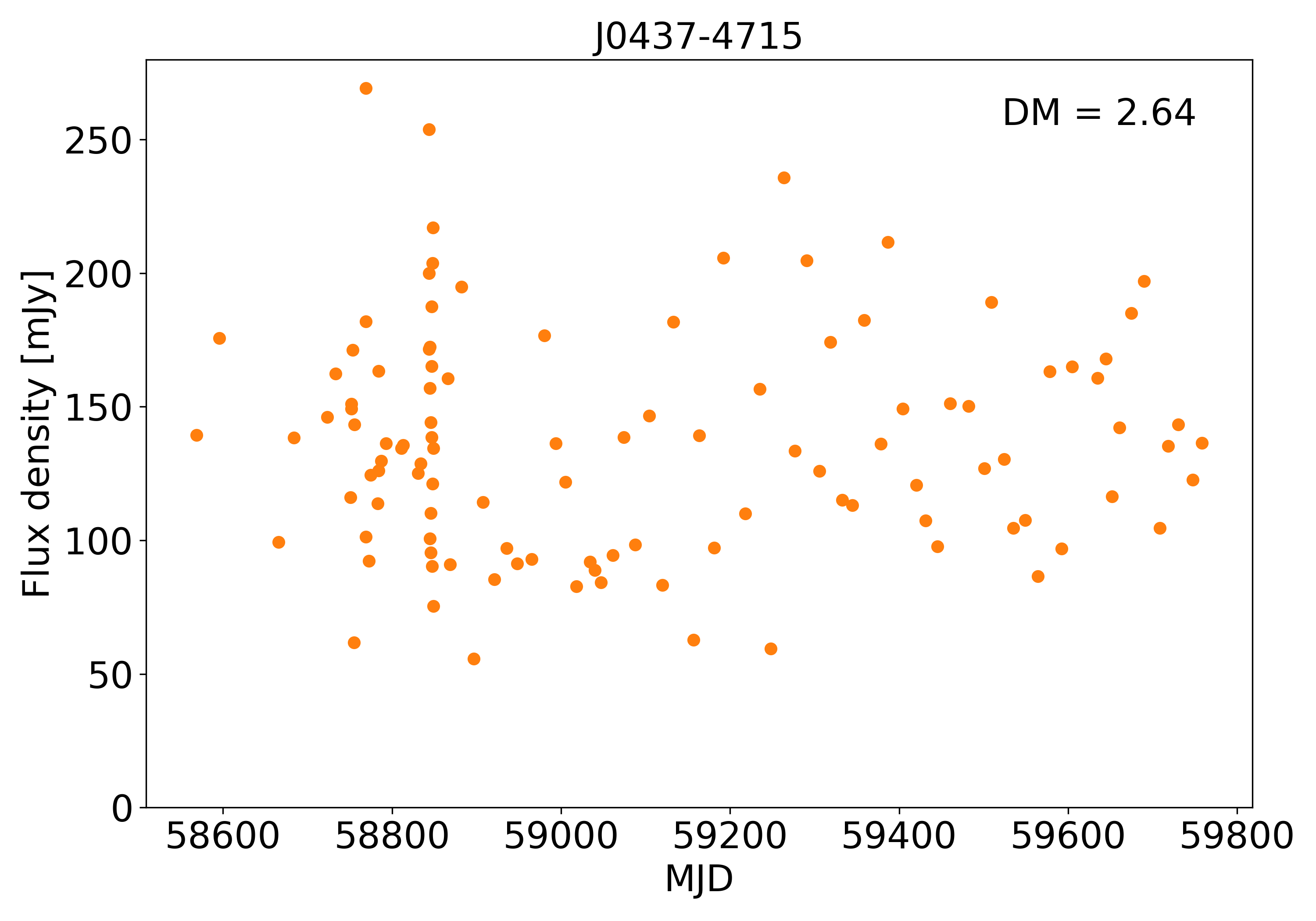}}
    \subcaptionbox{}
    {\includegraphics[width=0.47\textwidth]{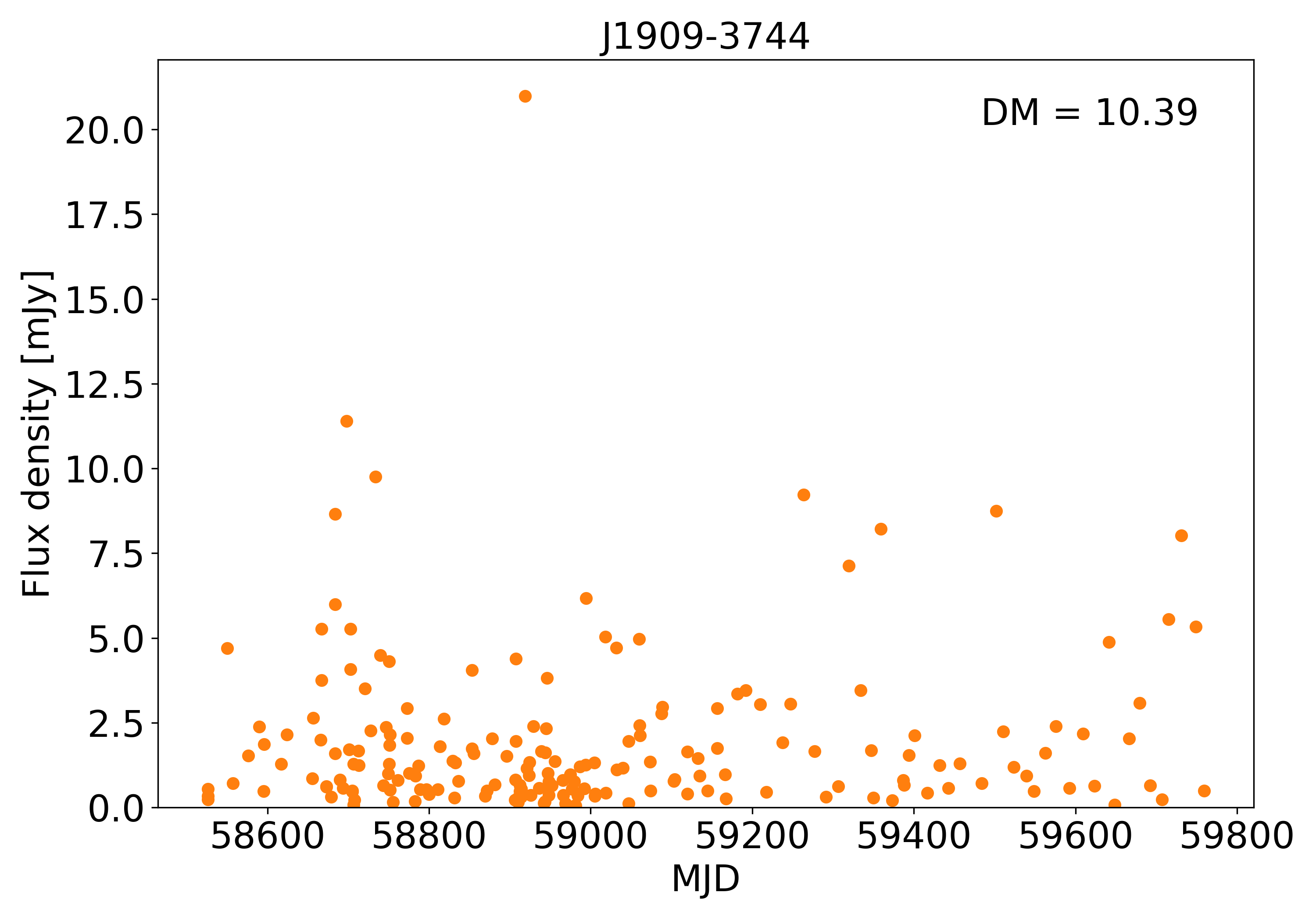}}\\
    \subcaptionbox{}
    {\includegraphics[width=0.47\textwidth]{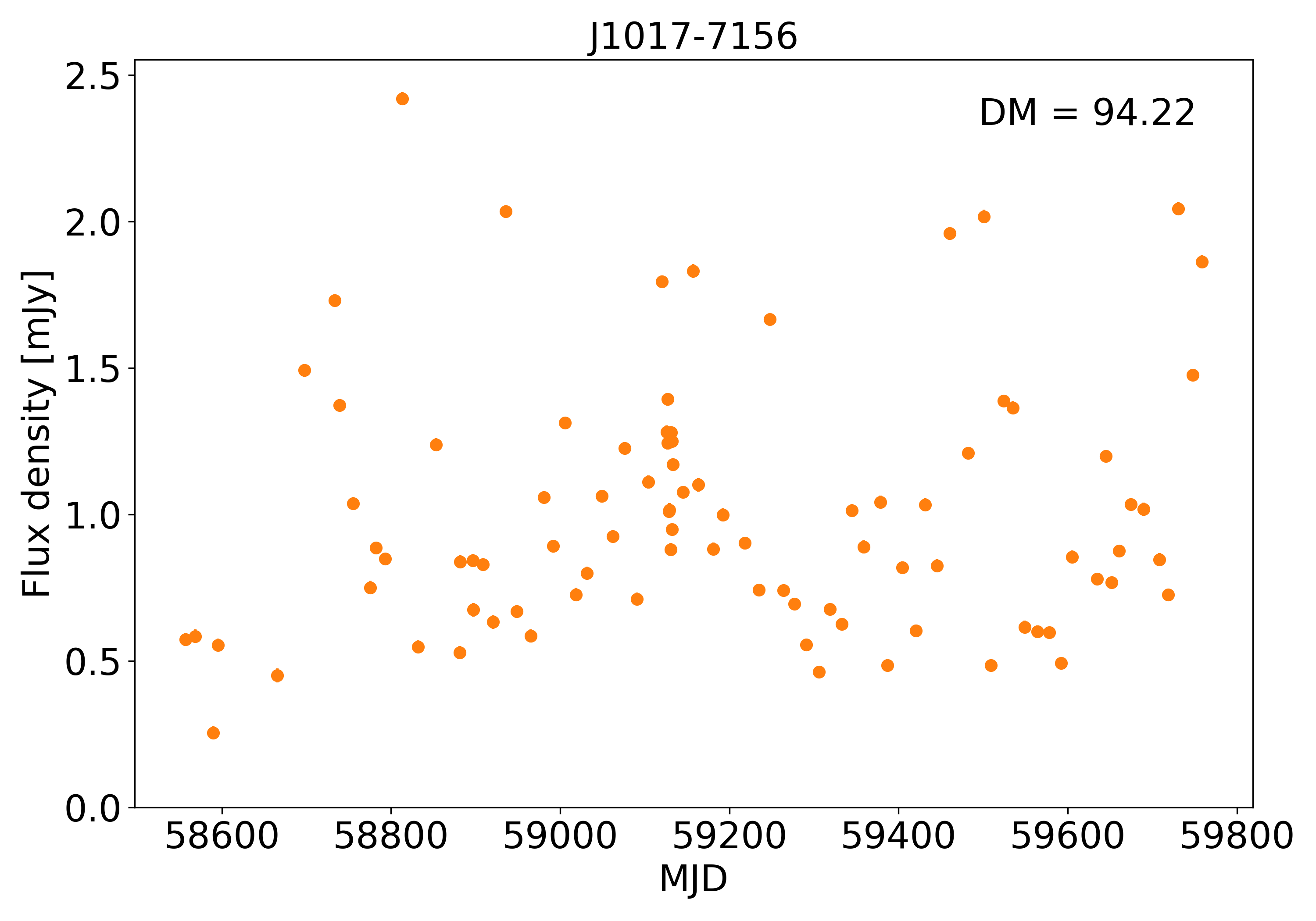}}
    \subcaptionbox{}
    {\includegraphics[width=0.47\textwidth]{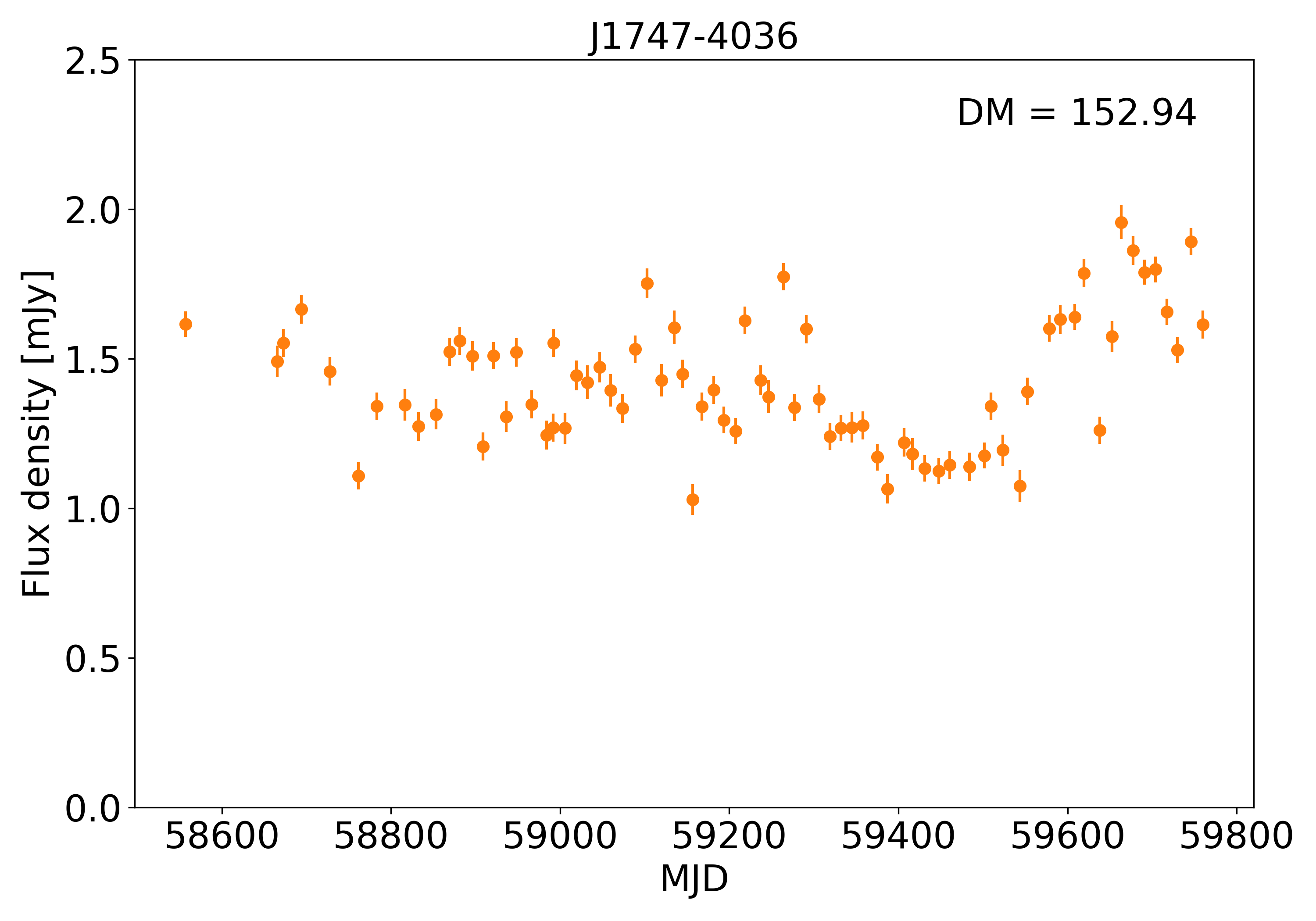}}\\

   \caption{Mean flux densities as a function of time for (a) PSR J0437$-$4715 (b) PSR J1909$-$3744 (c) PSR J1017$-$7156 (d) PSR J1747$-$4036 for a 97 MHz band at a centre frequency of 1429 MHz. Their modulation indices are (a) $0.3$ (b) $ 1.28$ (c) $0.43$ and (d) $0.15$. The error bars in the first three panels are too small to be distinguishable from the data points. The labels show their respective DMs in pc cm$^{-3}$. The apparent vertical stripes in panels (a) and (c) have arisen due to frequent observations of some pulsars as part of non-PTA observing campaigns aimed at studying orbital phase-dependent scintillation and Shapiro delays.}
\label{fig:mjdflx}
\end{figure*}

\section{Results} 
\label{sec:res}
In this section, we will first present the derived flux densities of our pulsar sample and 
then quantify their variability by calculating their modulation indices and temporal structure functions. Figure \ref{fig:psr_scint} shows the mean flux densities of four of
our pulsars (each of which has a very different DM) as a function of both pulse phase and radio frequency. The four pulsars were chosen
to show the different types of scintillation exhibited, ranging from
those with no extreme minima (except for those induced by interference) such as
J0437--4715, those with just one or two bright scintles present in the band like J1909--3744,
pulsars with many bright scintles across the band (J1017--7156) and those
with almost continuous intensity across the band like J1747--4036.
In Figure \ref{fig:mjdflx} we show how 
the flux densities of the same four pulsars 
change from epoch to epoch. Histograms of their
flux densities at a centre radio frequency of 1429 MHz are shown in
Figure \ref{fig:histogram}.


\subsection{Flux densities} 

\label{sec:flx_density}
Table \ref{tab:mod_ind} lists pulsar names, number of observations, DMs and mean pulsar flux densities measured at three different frequency bands centered at 944, 1429 and 1625 MHz of the
89 pulsars in our sample. We chose these three bands (of the 8) because
the 944 and 1625 MHz sub-bands
represent the extrema of our 775.75 MHz frequency band
and the 1400 MHz sub-band 
is the most often cited one in the literature for ease of comparison with other studies. See data availability statement for the values of these quantities at all 8 of the frequency sub-bands.
The large number of observations in the sample for each pulsar
(typically greater than 70) make it possible
to establish robust modulation indices (see Section \ref{sec:mod_ind}) for each of the pulsars. Another very useful quantity to demonstrate
the range of amplification experienced by the pulsars 
is the ratio of maximum flux density value observed and its median. 
This is a number rarely quantified in the 
literature but quite instructive when assessing the likelihood
that survey candidates should be re-observed in follow-up observations. 
At 1429 MHz, this ratio has a mean and median value of $\sim$5 and $\sim$3  respectively for our MSP sample. 
The maximum value of this ratio at 1429 MHz in our sample was $\sim$ 32
exhibited by PSR J2229+2643 which has a DM of 22.73 pc cm$^{-3}$.

To check that our method for determining the mean flux densities is accurate, we used the study of 
\cite{2015MNRAS.449.3223D} for comparison. Their study used a pulsed cal 
and observations of Hydra A at the 
Parkes 64\,m radio telescope to calibrate their flux densities. They had 21 sources in common with our sample
and the ratio of their flux densities to ours had a mean of 1.19 with a reasonably high standard deviation of 0.2. Unfortunately many of their pulsars were part of timing array observations
where the observer has discretion to alter the observing schedule. Observers could choose to
abandon observations of pulsars that were in a low-flux state,
or repeat observations of MSPs experiencing a scintillation maximum, potentially
biasing their sample.
The highest DM pulsar that was common between our samples was J1017--7156, 
which agreed with our flux density to within 1\% and has a DM of 94.22 pc cm$^{-3}$. 
Another flux density study by \cite{2023MNRAS.tmp..221W} studied 28 pulsars in common with ours, 
and with 15 possessing DMs greater than 60 pc cm$^{-3}$. 
The average ratio of their mean flux densities to ours
was 0.84 
with a standard deviation of 0.19. Clearly, our flux densities are
comparable to those of other authors, but may have a bias
if the supplied system equivalent flux density from SARAO
is in error, or our assumptions about RFI cleaning are invalid. From our comparisons above we suspect that any systematic bias is at most 20\%.


\subsection{Spectral Indices} 
\label{sec:spec_ind}
The radio spectra of pulsars 
as a function of frequency are often well modelled as a power law
\begin{align}
    \label{eq:spectra}
    S = S_{0} 
 \bigg(\dfrac{\nu}{\nu_{0}}\bigg)^{\alpha},
\end{align}
where $ S_{0}$ is the flux density at the reference frequency $ \nu_{0}$ and $\alpha$ is the spectral index. 
In our study, the spectral index for each pulsar was calculated using the mean flux density values at each frequency sub-band and fitting a simple 
power law via a least-squares method.

\begin{figure}
\centering
\includegraphics[width=0.49\textwidth]{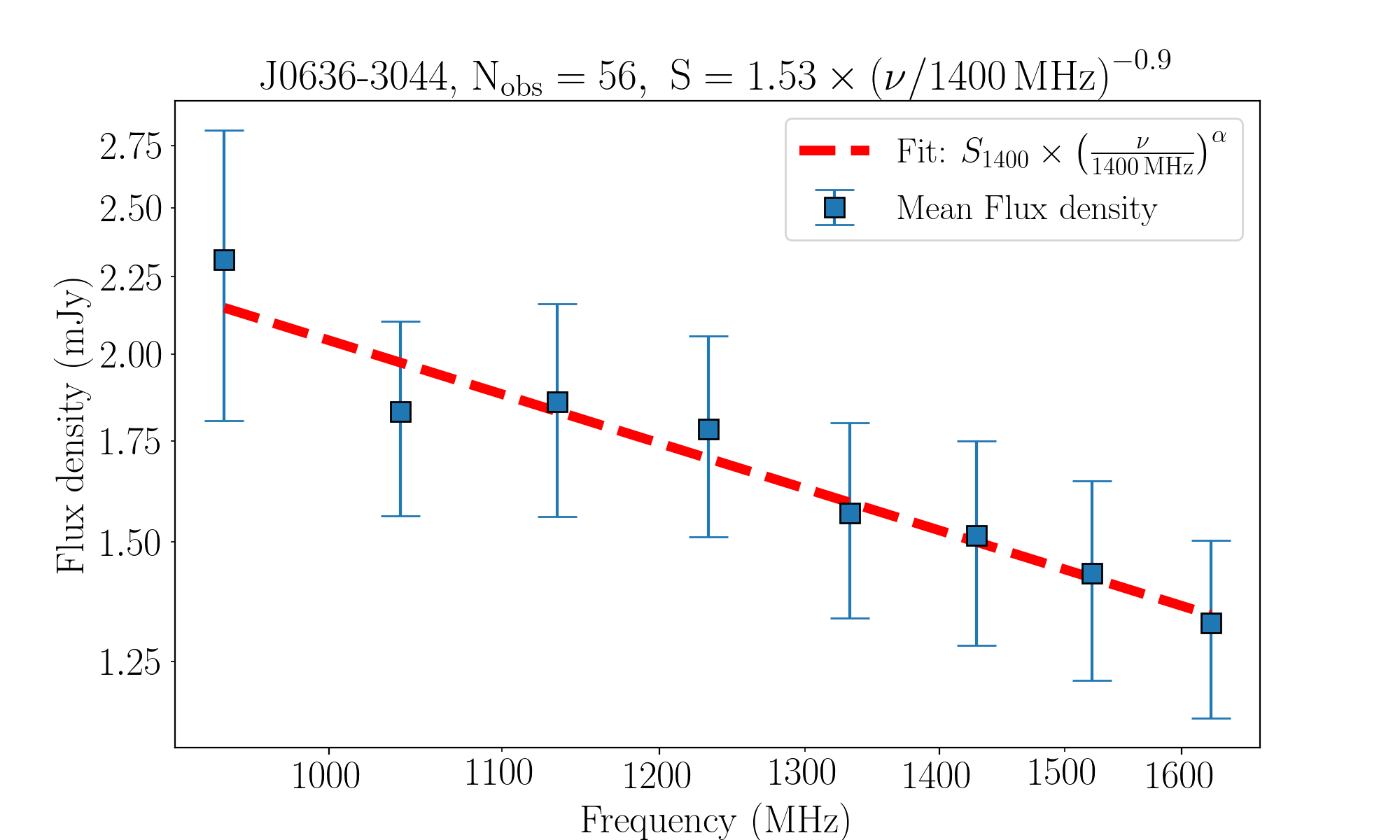}
\includegraphics[width=0.49\textwidth]{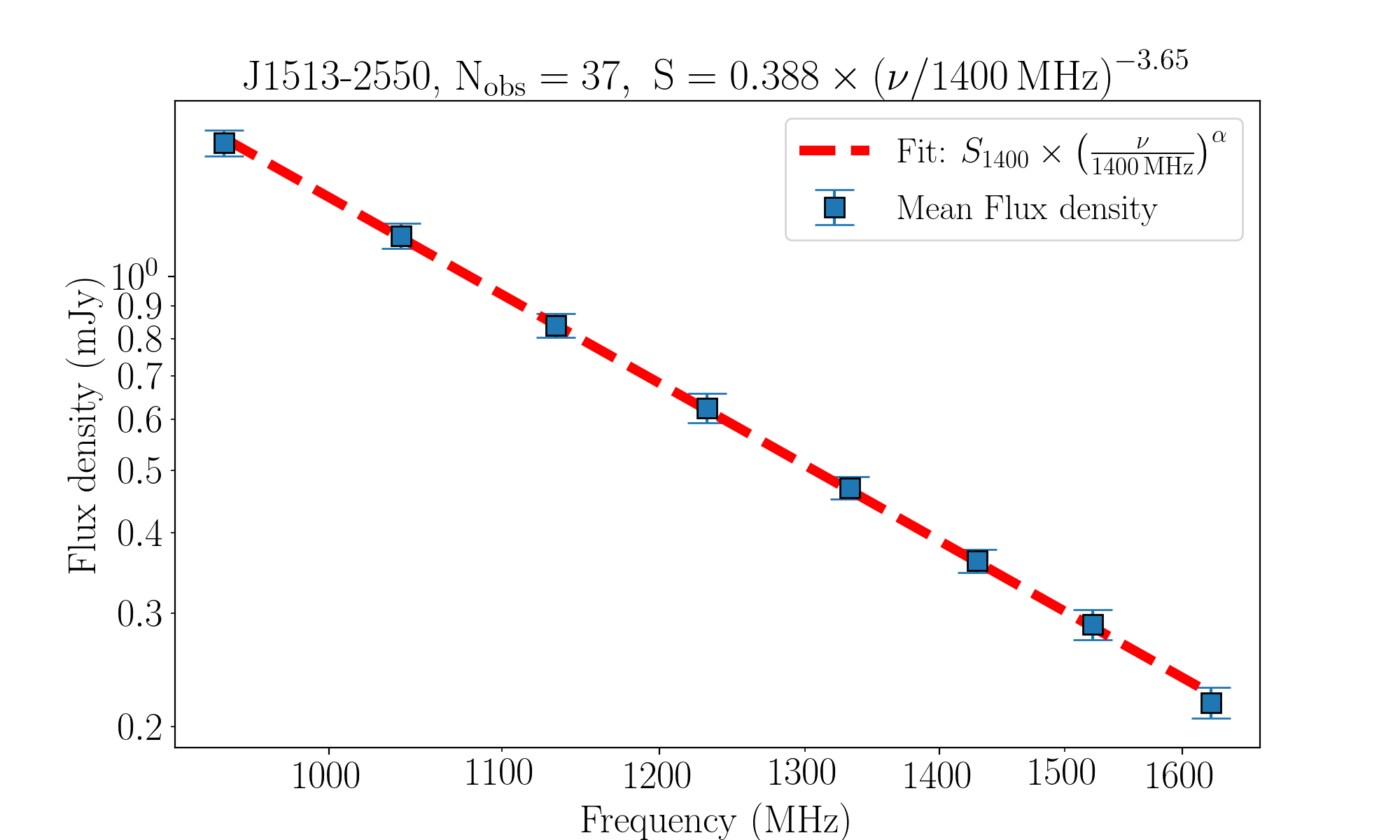}
\includegraphics[width=0.49\textwidth]{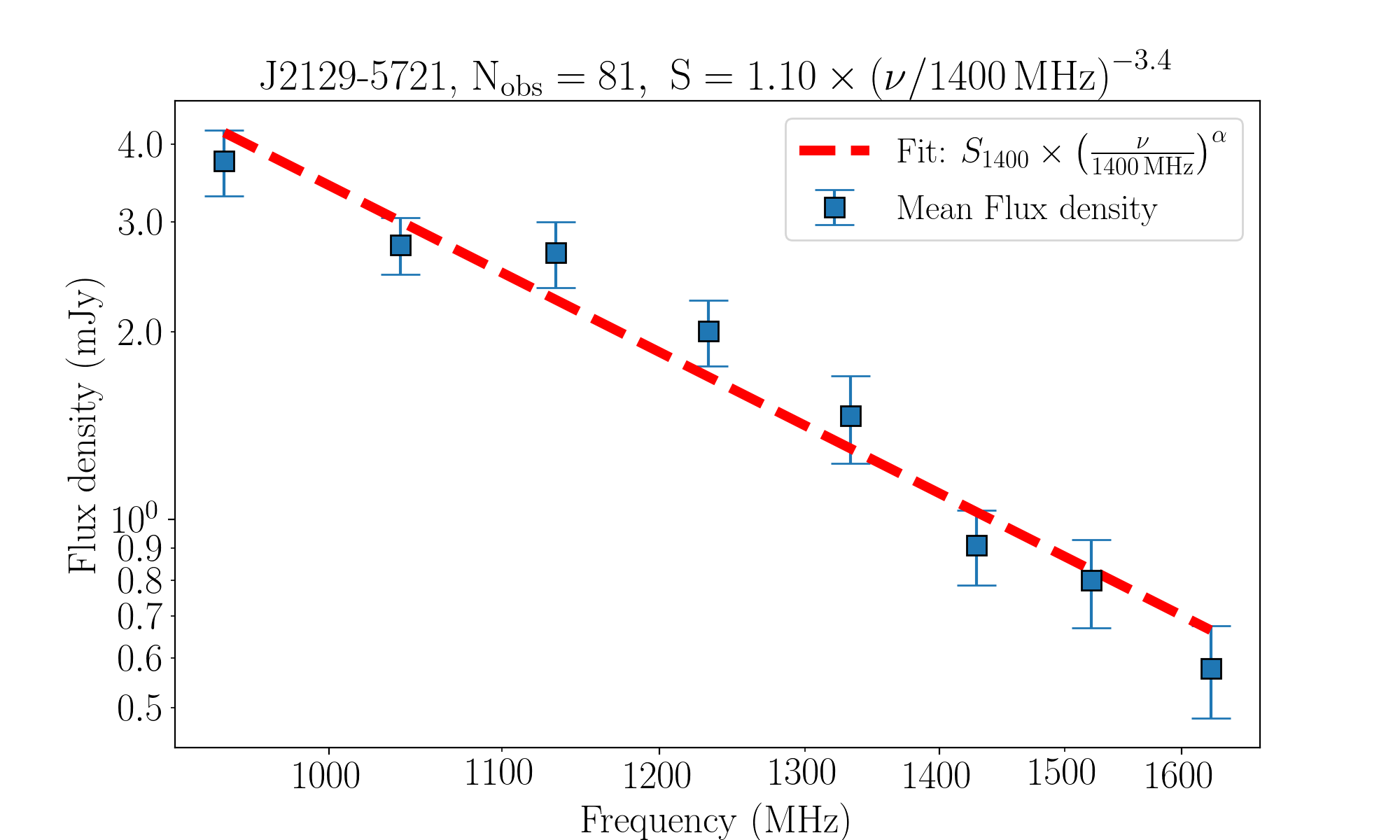}
\caption{Spectra of PSR J0636--3044 (the flattest in our sample) and PSR J1513--2550 (steepest) and J2129--5721, demonstrating the range of pulsar flux density spectra from our sample.}
\label{fig:msp_spec}
\end{figure}

The last two columns in Table \ref{tab:mod_ind} list the spectral index and its error calculated using the mean flux densities in the 8 observing frequency sub-bands and a least squares power law fit and Figure \ref{fig:msp_spec} shows spectral index fit for 3 MSPs in our sample.
All of the MSPs in our sample have negative spectral indices ranging from $-0.9(3)$ for PSR J0636--3044 based on 56 epochs to $-3.65(10)$
for PSR J1513--2550 which has 37 observations. The overall sample has a mean spectral index of $-1.86(6)$. This result is consistent with the findings from \cite{1998ApJ...506..863T} who found a mean value of $-1.8$.  For the 21 MSPs in common between our sample and \cite{2015MNRAS.449.3223D}, we find a mean spectral index value of $-1.89(13)$ whereas they find $-1.71(8)$. Since their lower frequency band extends down to $\sim$700 MHz, this may be
due to a slight flattening of the spectra at lower frequencies as is clearly exhibited by
PSRs J1600--3053 and J1713+0747 in their sample. The biggest discrepancy between the spectral indices of our data sets was exhibited by PSR J2129--5721 where we measure a spectral index of $-3.4(3)$ as compared to their much flatter $-2.12(7)$. This pulsar has a modulation index of $1.2$ at 1429 MHz and hence the mean flux density can vary substantially. From the spectral index fit plot given in their paper for this pulsar, it is also evident that their observations show a large step-change flux density deviation from a linear trend at 3 GHz, probably also due to scintillation. We are therefore not unduly concerned by
our slightly steeper spectra.

 To determine how well the power law describes our pulsar spectra, we performed a chi-squared test. 
 For most of the pulsars, the reduced $\chi^2$ value was near unity with only two notable exceptions. 
 PSR J1804$-$2858 and PSR J1756$-$2251 both deviated quite strongly from the single power law fit.
 In the case of PSR J1756--2251 the spectrum appears curved, with evidence of mild flattening
 at lower frequencies. PSR J1804--2858 on the other hand has the highest DM of all the pulsars
 in our sample with DM = 232.5 pc cm$^{-3}$, and shows evidence of a scattering tail that spans
 the entire pulse period in the lowest frequency channels. This negates our assumption
 in Eqn \ref{eq:flux} that there exists a baseline that contains none of the pulsar's flux density. At low frequencies some of the pulsar's flux density is thus absorbed into the baseline 
 leading to an apparent spectral turnover where none might actually exist.
 For the majority of our pulsars however, there is no evidence for an appreciable spectral turnover over our frequency range. 
For the population as a whole we investigated whether the average residuals
from the power law fit showed any structure as a function of frequency 
by determining the ratio of the mean flux density to the power law fit 
in each channel, and then averaging these over the entire population.
We found that the mean deviation from the average power law was at most 3\%
with the lowest frequency bins 1.5\% lower, the central bins a few percent
higher, and the highest bin lower by 2\% suggesting the population
as a whole might experience a very mild spectral turnover at the frequencies observed.


\subsection{Flux density Modulation Indices} 
\label{sec:mod_ind}

To quantify the variability of flux densities of pulsars, a useful parameter 
is the modulation index
\begin{align}
\label{eq:mod_index}
 m = \dfrac{\sigma_{\rm S}}{S_{\rm mean}},
\end{align}
\noindent where $\sigma_{S}$ is the standard deviation of the flux densities and $ S_{\rm mean}$ is the mean flux density in an observing frequency sub-band.
Inspection of Table \ref{tab:mod_ind} reveals that the modulation indices of the pulsars depend strongly upon the DM of the pulsar. Higher DM pulsars have a lower modulation index value in comparison to low DM pulsars in general. From Table \ref{tab:mod_ind} we find that our largest modulation index was 2.0(4)
for the DM = $10.39$ pc cm$^{-3}$ pulsar J1909--3744 at a centre frequency of 1625 MHz based upon 192 observations, while 
the lowest modulation index of 0.08(1) was for the DM = $114.54$ pc cm$^{-3}$ pulsar
J1843--1448 at a centre frequency of 944 MHz based upon 38 observations.  
There is also a tendency for intermediate DM pulsars to exhibit higher modulation indices at higher frequencies.
At 944 MHz, the mean modulation index of the sample is 0.56 and median is 0.40, with a standard deviation of the population 0.39. As we go to higher centre frequencies, at 1429 MHz, the mean, median and standard deviation of modulation index is 0.63, 0.50 and 0.41 respectively.
At a centre frequency of 1625 MHz, the mean, median and standard deviation shifts to 0.68, 0.60, and 0.44 respectively. At higher radio frequencies, the differential path lengths of scattered waves are shorter, which results in larger decorrelation scales with frequency and time, and greater modulation of the flux densities.

\begin{figure}
\centering
\includegraphics[width=0.48\textwidth]{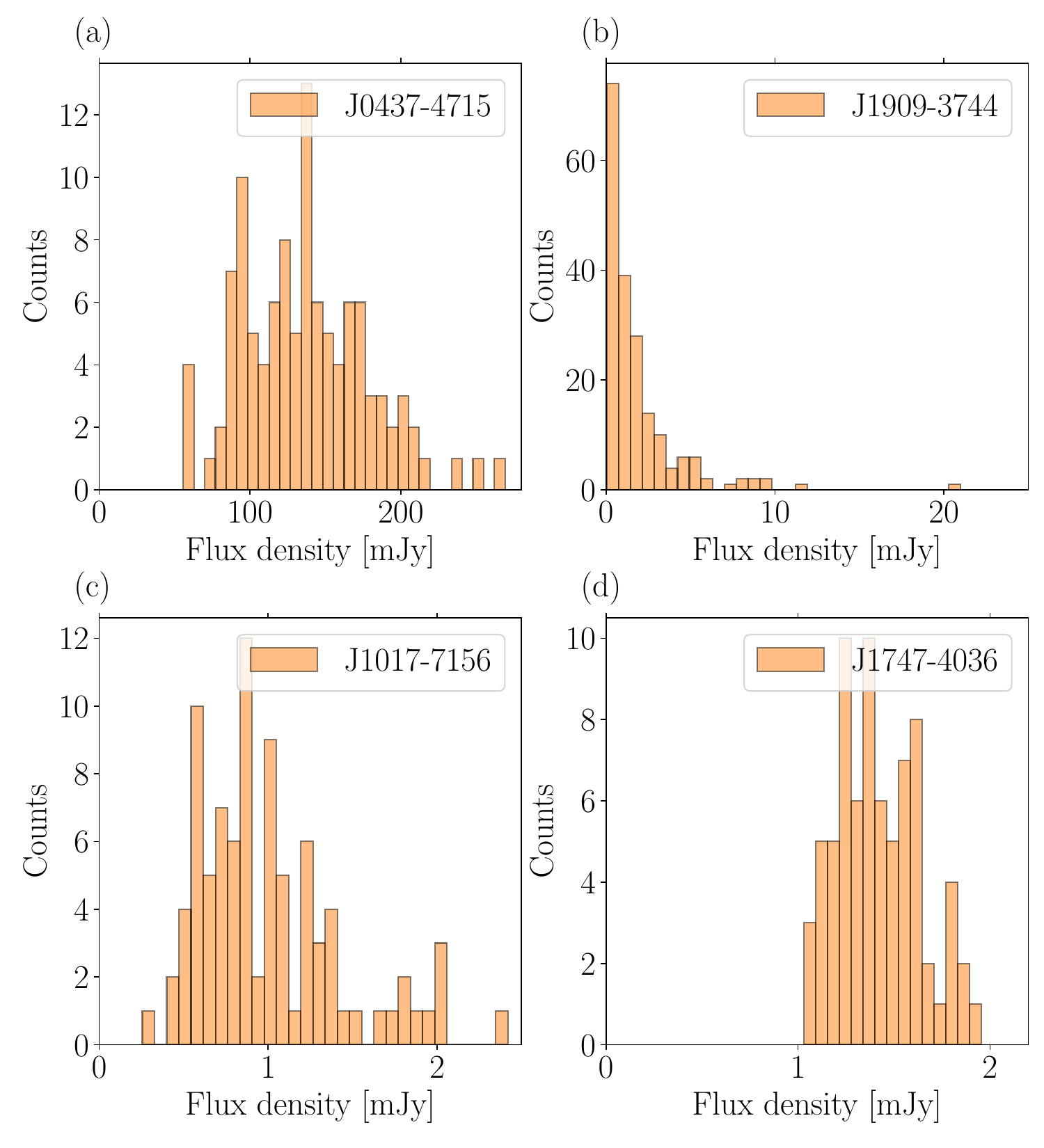}

\caption{ Flux density histograms of four pulsars at a centre frequency of 1429 MHz. PSR J0437--4715 with DM 2.64 pc cm$^{-3}$, PSR J1909--3744 with DM 10.39 pc cm$^{-3}$, 
 PSR J1017--7156 with DM 94.21 pc cm$^{-3}$ and PSR J1747--4036 with DM 153 pc cm$^{-3}$. As the DM changes we see evolution from
a Gaussian-like flux density distribution for J0437--4715 with a small exponential tail, to an almost pure
exponential for J1909--3744, back to a Gaussian for J1017--7156 that becomes even more
centrally condensed for the high DM pulsar J1747--4036.}
\label{fig:histogram}
\end{figure}


\subsection{Flux Density Structure Functions} 
\label{sec:stru_func}
The pulsar flux densities vary on a large range of timescales.
A common way to characterise such changes as a function
of time is via the use of a temporal structure function. 
In this way, we can evaluate relative changes on timescales of weeks to years.
The flux densities of four pulsars at different DMs are plotted in Figure \ref{fig:mjdflx} that illustrate how different pulsar 
flux densities change with time. Depending on the DM of the pulsar, the flux density either changes independently between observations as shown for pulsars J0437$-$4715, J1909$-$3744 and J1017--7156 
or more slowly for J1747$-$4036.
We attribute the slow changes in flux density for J1747--4036 are caused by refractive scintillation.
To quantify the effects of refractive scintillation on pulsar flux
densities, the structure function of the time series of each pulsar was calculated. The structure function formula used here follows the definition by \cite{2021MNRAS.501.4490K} and gives a measure of
the relative change as a function of time. Small values ($\ll 1$) indicate 
low fractional changes, whereas high values point to large variations.
In our study the structure function was calculated in bins of lag starting from 10 days to around 1200 days since our total data span is at
most $\sim$3 years.

\noindent The structure function for a time lag of $\tau $ is given by
\begin{multline}
     D(\tau) = \dfrac{1}{\Bar{S}^{2} N_{\rm p}}  \Biggl\{ \sum_{ij}^{ N_{\rm p}} (S_{\rm i} - S_{\rm j})^2 + \sum_{ij}^{N_{\rm p}} (e_{\rm i}^2 +e_{\rm j}^2) \\  + 2 \sum_{ij}^{N_{\rm p}} ( e_{\rm i} - e_{\rm j}) (S_{\rm i} - S_{\rm j}) - 2 \sum_{ij}^{N_{\rm p}} e_{\rm i} e_{\rm j} \Biggr\},  
\end{multline}

\noindent where $\Bar{S}$ is the average flux density, $S_{\rm i}$ is the individual flux densities at each epoch, $N_{p}$ is the number of pairs of flux density points that are separated by this lag and $e_{\rm i}$ is the uncertainty in measurement of the flux density values.
\noindent The error in the calculations of the structure functions can be estimated 
following the derivation in appendix A of \cite{2007MNRAS.378..493Y} as

\begin{equation*}
      {\sigma_{D}}^{2} (\tau) =  \dfrac{1}{N_{p}^{2}}  \Biggl\{ \sum_{i}  N_{\rm i}^2 e_{\rm i}^4 + 4 \sum_{i}\sum_{j} e_{\rm i}^2 (S_{\rm i} -  S_{\rm j})^2  + 4 \sum_{ij} e_{\rm i}^{2} e_{\rm j}^{2} \Biggr\},   
\end{equation*}

\noindent where $N_{i}$ is the number of times a flux density value is used to calculate the structure function at a particular lag.

Structure functions are widely used to study time series and scintillation effects are usually unbiased by irregularly sampled data and hence ideal for studying the refractive scintillation in temporal flux density observations. The shape of the function can also be used to estimate the refractive scintillation timescale. These parameters derived from the structure functions inform us about the scattering material or inhomogeneities along the line of sight to the pulsars. 

A typical structure function has three distinct regions. At very small lags, the structure function is flat and sensitive to calibration noise, radiometer noise, diffractive noise and pulse to pulse variations introduced due to intrinsic variability of pulsars. 
This is also known as the ``noise'' regime. We do not probe this region in our study.
At medium lags ($>$ days), the structure function has an increasing slope and can be modelled using a simple power law on a log-log scale. The log-log slope is usually less than 1 as shown by \cite{1990ApJ...352..207S} and \cite{2021MNRAS.501.4490K}. 
At large lags in excess of the refractive timescale, the structure function saturates and hence flattens. The lag at which the structure function reaches half of its saturation value is often defined as the refractive scintillation timescale \citep{2000ApJ...539..300S}. 

Table \ref{tab:stru_func} shows the structure function values of all the 89 MSPs in our sample at lags of 2 weeks, 1 month, 1 year and 3 years with errors associated with them in brackets. The structure functions shows the complex nature of the temporal variations of the MSPs but can
be used to help plan observations, especially if recent
observations of the pulsar exist and it has a long refractive
timescale. The extremely low values of the 2-week
structure function for some pulsars (e.g. J1643--1224 with 0.008(5))
give us great confidence in the stability of the MeerKAT phasing and
calibration, and also the validity of our flux density calibration procedure at least on a relative scale.

\begin{figure*}
\centering
    \subcaptionbox{}
    {\includegraphics[width=0.48\textwidth, scale=1.7]{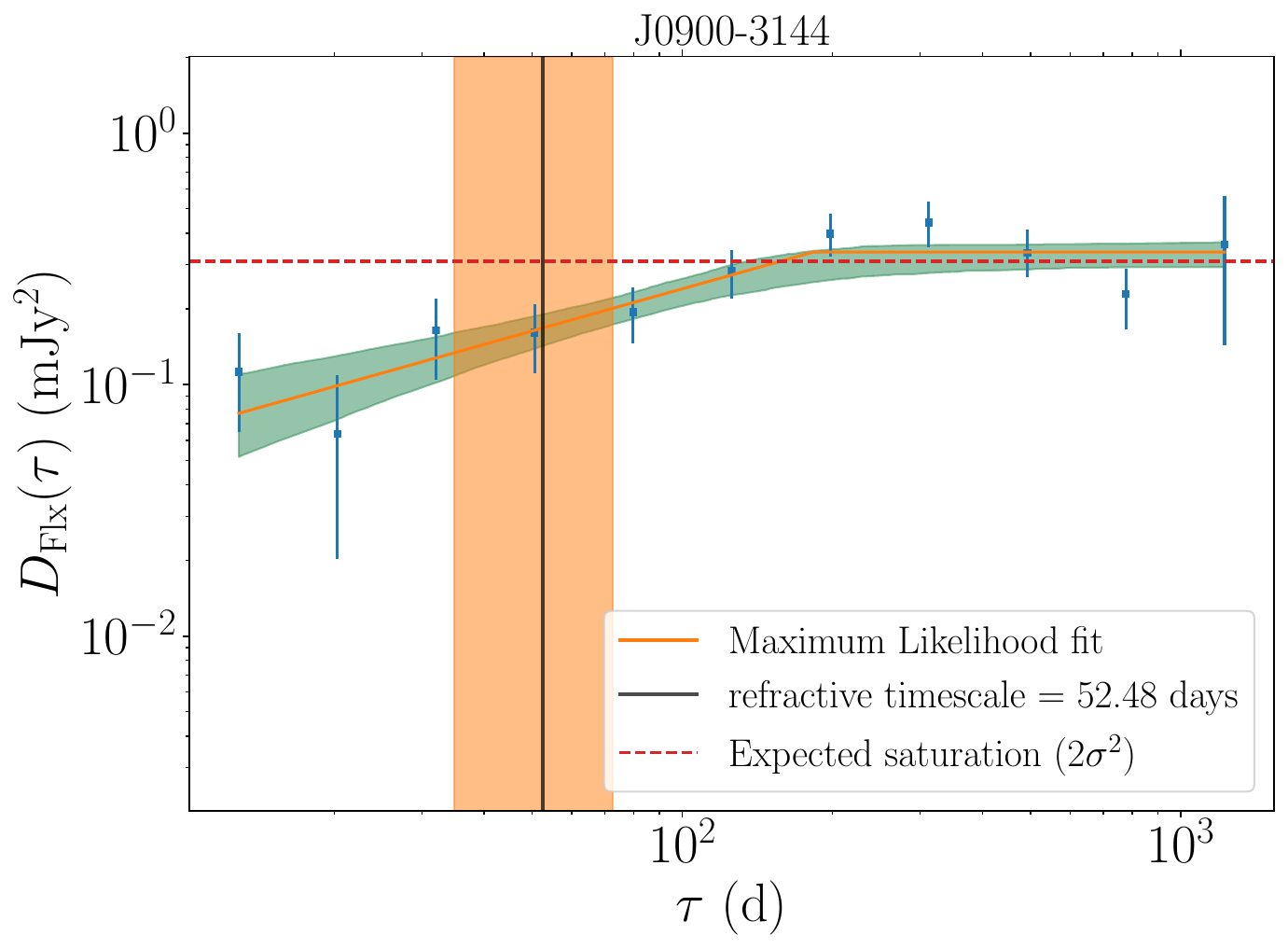}}
    \subcaptionbox{}
    {\includegraphics[width=0.48\textwidth, scale=1.7]{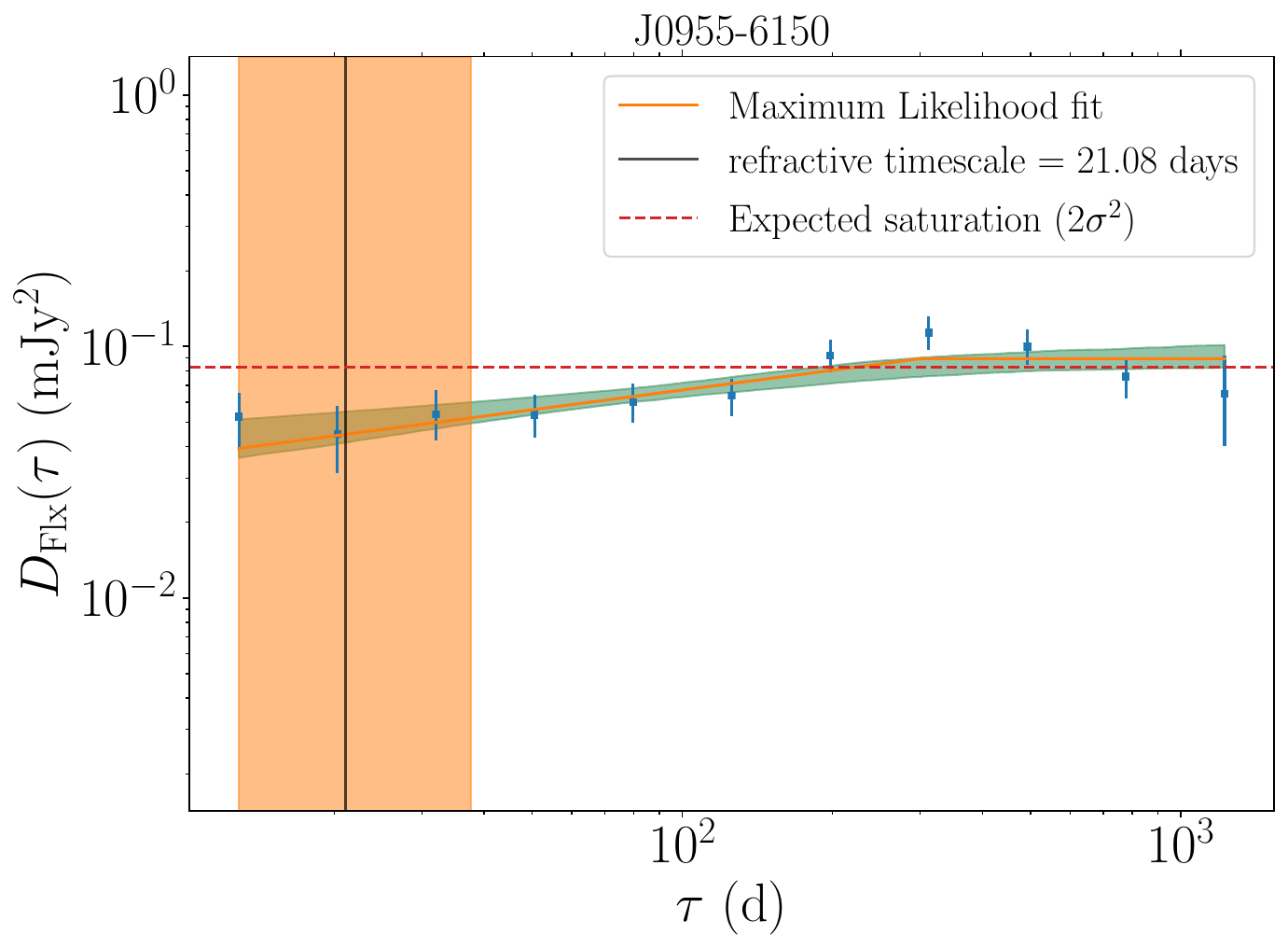}}\\
    \subcaptionbox{}
    {\includegraphics[width=0.48\textwidth, scale=1.7]{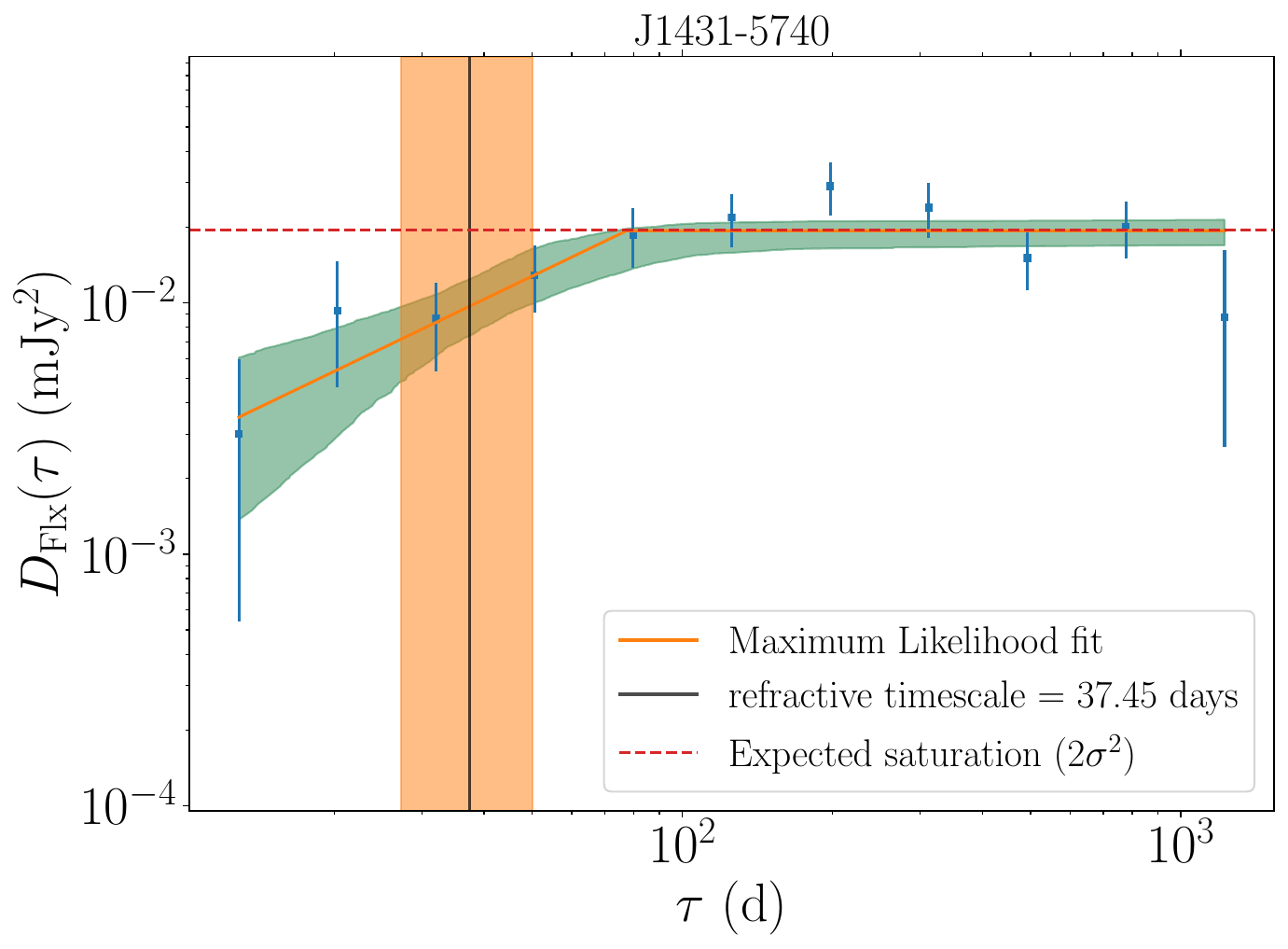}}
    \subcaptionbox{}
    {\includegraphics[width=0.48\textwidth, scale=1.7]{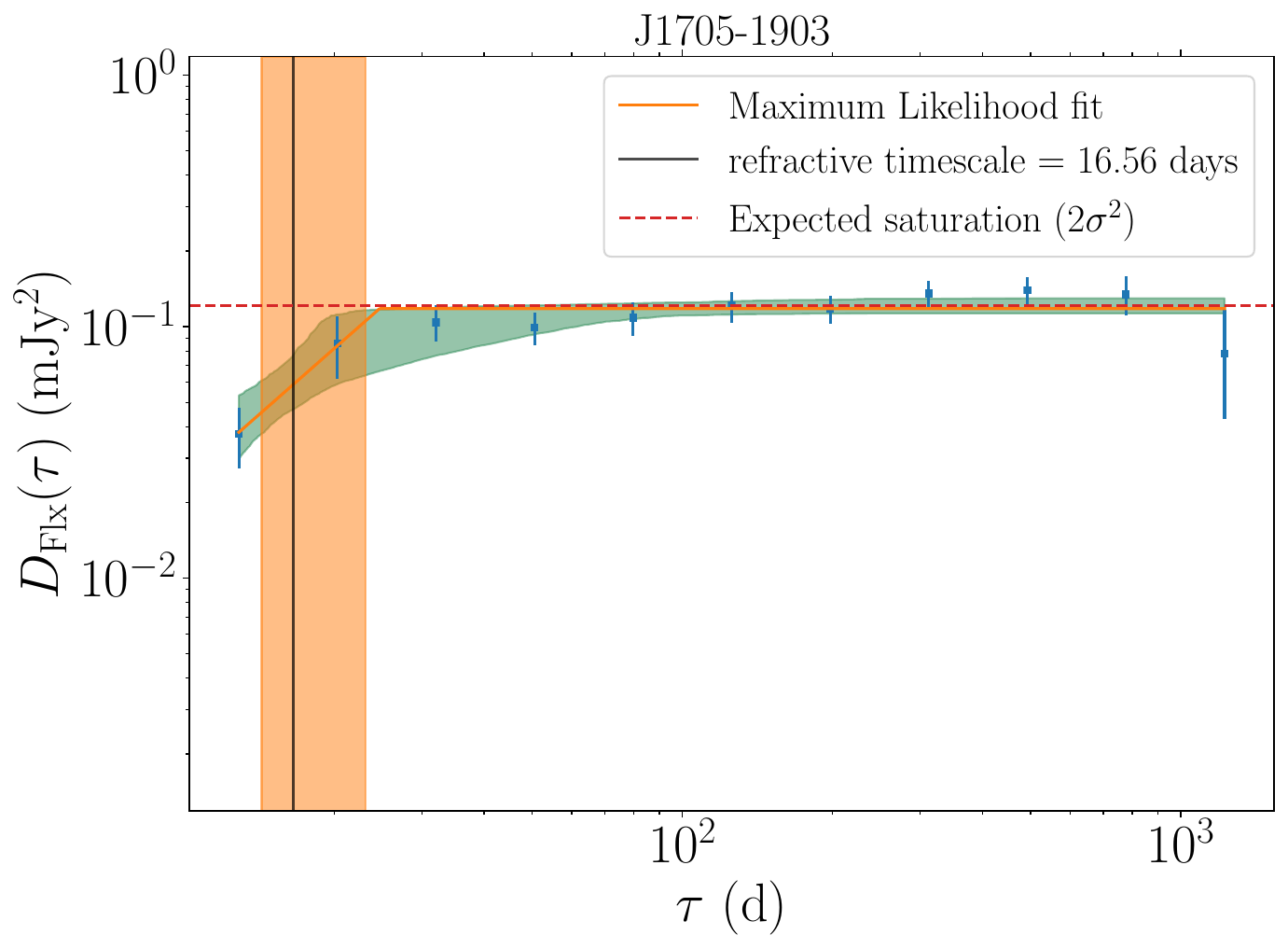}}\\

    \subcaptionbox{}
    {\includegraphics[width=0.48\textwidth, scale=1.7]{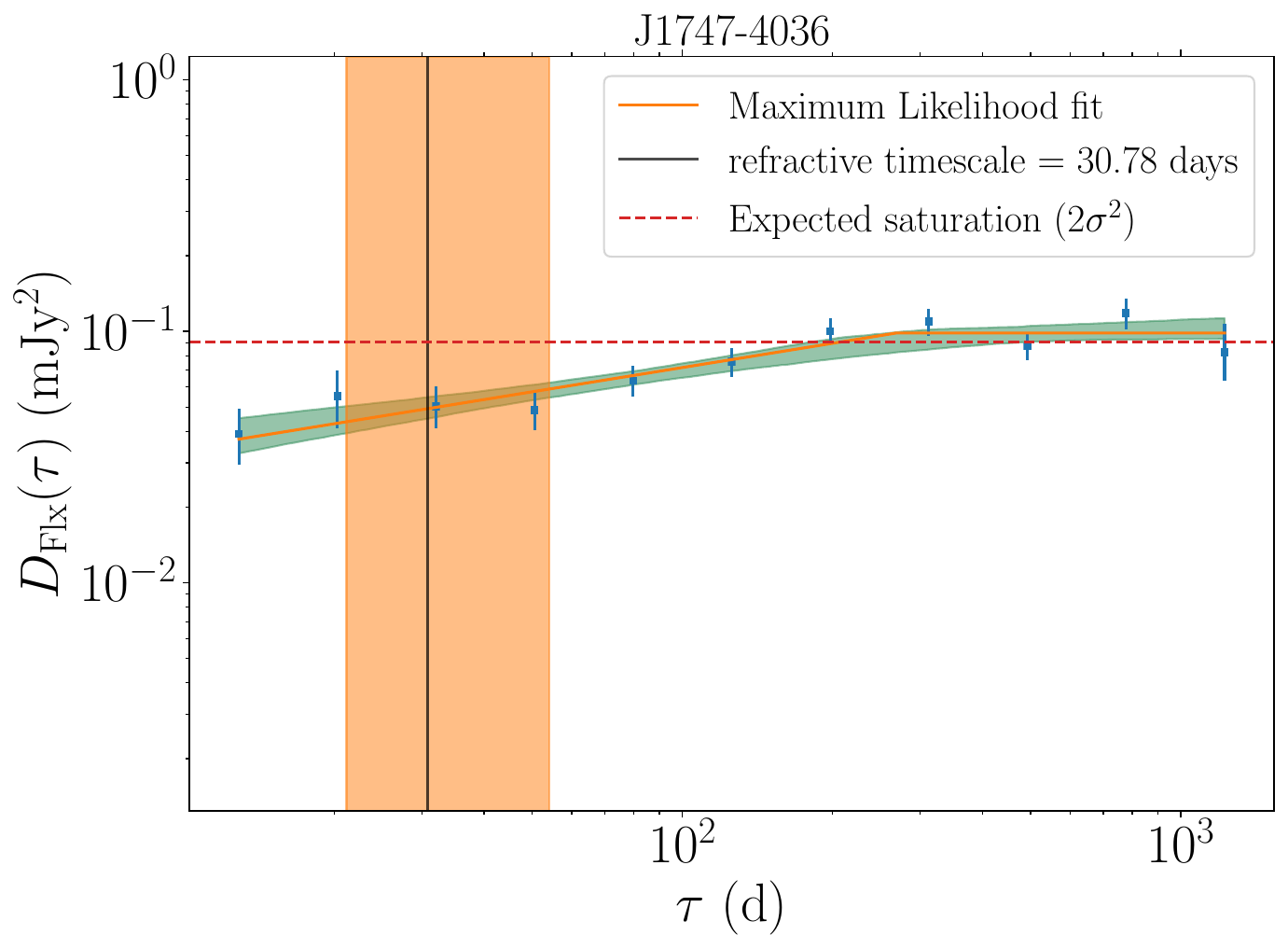}}
    \subcaptionbox{}
    {\includegraphics[width=0.48\textwidth, scale=1.7]{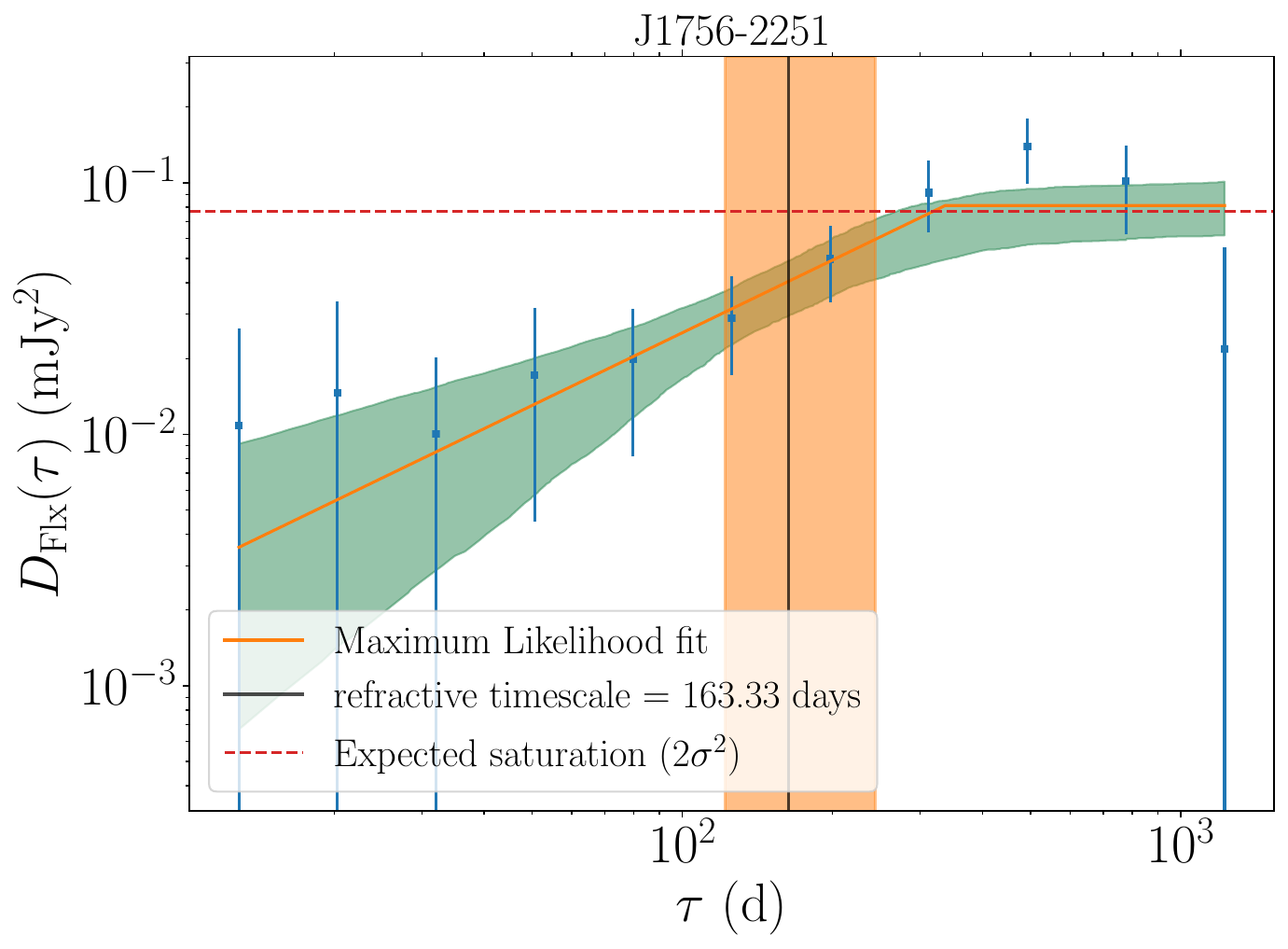}}\\
\end{figure*}
\begin{figure*}\ContinuedFloat
    \centering
    \subcaptionbox{}
    {\includegraphics[width=0.48\textwidth, scale=1.7]{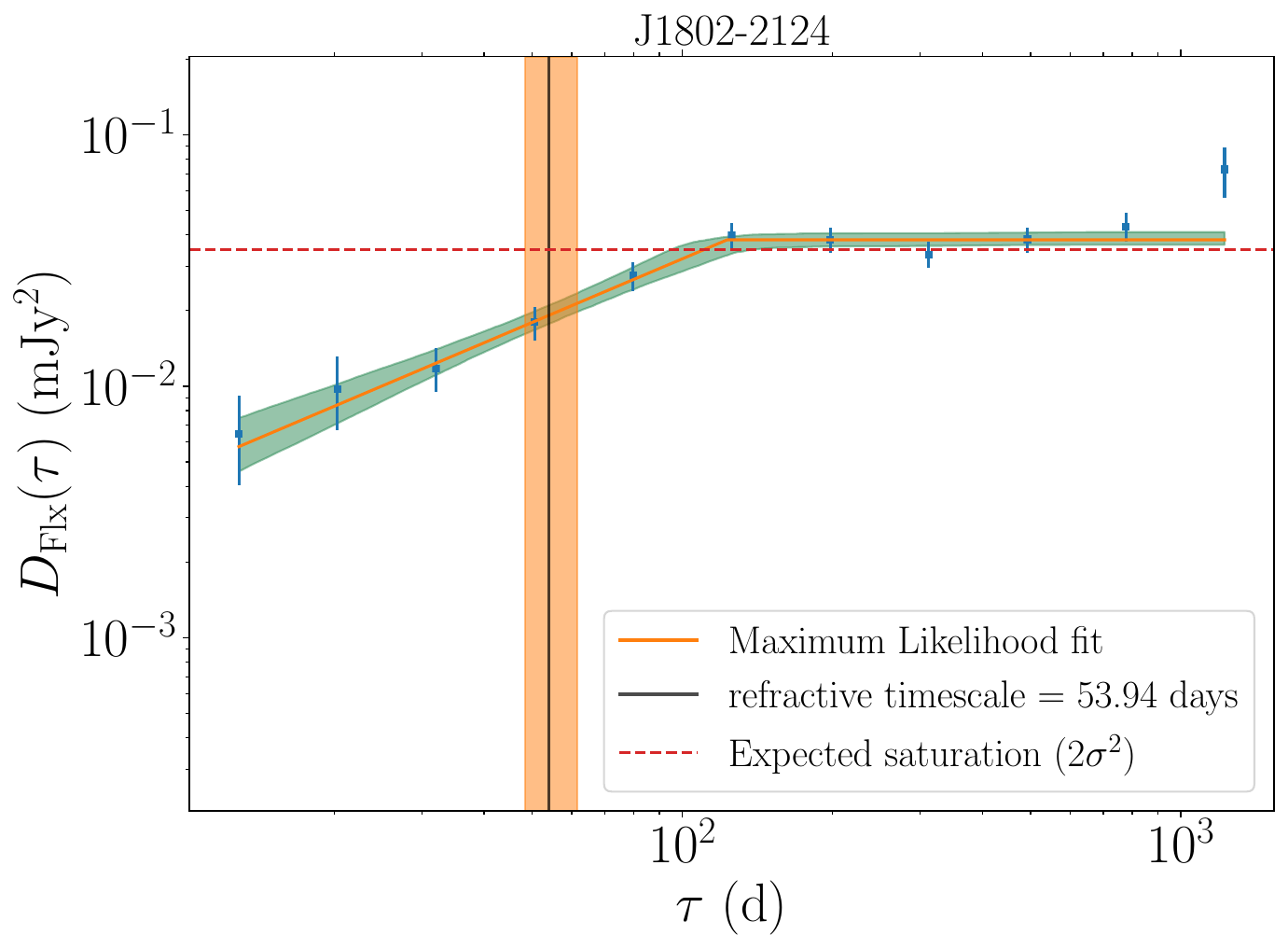}}

\caption{Structure function as a function of time lag for (a) PSR J0900--3144 (DM = 75.7 pc cm$^{-3}$) (b) PSR J0955--6150 (DM = 160.9 pc cm$^{-3}$) (c) PSR J1431--5740 (DM = 131.4 pc cm$^{-3}$) (d) PSR J1705--1903 (DM = 57.5 pc cm$^{-3}$) (e) PSR J1747--4036 (DM = 152.9 pc cm$^{-3}$) (f) PSR J1756--2251 (DM = 121.2 pc cm$^{-3}$) (g) PSR J1802--2124 (DM = 149.6 pc cm$^{-3}$) for a 97 MHz band at a centre frequency of 1429 MHz. The shaded orange and green regions show the error associated with the measurement of the refractive timescale and the maximum likelihood fit for the structure function respectively. The red dotted line shows the expected saturation level for a random process with variance $\sigma^{2}$ for time lags much larger than the associated timescale. }
\label{fig:refract_scint}
    
\end{figure*}


There are a few different classes of structure functions.
Some pulsars show linear increase in the structure 
function before reaching saturation. Other - mainly low
DM pulsars - show no temporal evolution and are essentially constant
and cannot have their refractive timescales estimated.
Some pulsars only have the increasing slope region and insufficient
time spans to reveal the flattened region.
In order to estimate the refractive scintillation timescale for our MSPs, we plotted the structure function as function of time lags using the equations described in \cite{2004ApJ...609..776H} using the publicly available structure function package from Thomson\footnote{https://github.com/AlecThomson/structurefunction}. Figure \ref{fig:refract_scint} shows the computed structure function of 7 MSPs at a central frequency of 1429 MHz. 
Only seven of our pulsars exhibited linear trends in their log-log plots
and clear saturation.
The refractive scintillation timescale for the 7 pulsars for which we could
 measure it are presented in Table \ref{tab:refract_scint}. 

\begin{table}
\centering
\caption{Table showing the measured refractive scintillation timescale and the slope of structure function of 7 pulsars in our sample of MSPs.}
\label{tab:refract_scint}
\begin{tabular}{rrr}
\multicolumn{1}{c}{NAME}  &  \multicolumn{1}{c}{$\tau_{\rm r}$}  & \multicolumn{1}{c}{Slope}\\
& (days)  & \\
\hline
\vspace{1mm}
J0900--3144 & $52^{+20}_{-17}$ & $0.53^{+0.21}_{-0.20}$\\
\vspace{1mm}
J0955--6150 & $21^{+16}_{-8}$  & $0.21^{+0.08}_{-0.07}$\\
\vspace{1mm}
J1431--5740 & $37^{+13}_{-10}$  &  $0.89^{+0.89}_{-0.38}$\\
\vspace{1mm}
J1705--1903 &$16^{+6}_{-2}$ & $1.14^{+2.06}_{-0.69}$\\ 
\vspace{1mm}
J1747--4036 &$31^{+23}_{-9}$ & $0.29^{+0.09}_{-0.08}$\\
\vspace{1mm}
J1756--2251 & $163^{+80}_{-41}$ & $0.91^{+0.75}_{-0.35}$\\
\vspace{1mm}
J1802--2124 & $54^{+8}_{-6}$ & $0.84^{+0.16}_{-0.16}$    \\    
\hline   
\end{tabular}
\end{table}

Seven MSPs in our sample are in common with the study of \cite{2023MNRAS.tmp..221W}, 
so we compared our results. Only two of the MSPs had refractive timescales ($\tau_{\rm r}$)
measured in both studies but they are both in good
agreement. For PSR J0900--3144 we measured $\tau_{\rm r} = 52^{+20}_{-17} $\,d whereas they
found $\tau_{\rm r} = 51.7$\,d, while for J1802--2124,
we had $\tau_{\rm r} = 54^{+8}_{-6} $\,d and they had $\tau_{\rm r} = 50.5$\,d.

The slope of the structure function in the log-log plot
provides information about the power spectrum of the
inhomogenities in the ISM. 
Most of the pulsars have a slope < 1 which is consistent with the studies of \cite{2021MNRAS.501.4490K}. The slope of the structure function is related to the IISM distribution along the line of sight. As suggested by \cite{1986MNRAS.220...19R} 
a slope of $\sim$ 2 would be indicative of a Kolmogorov spectrum with a thin screen model, but in our sample, only the intermediate-DM MSP J1705--1903
had a slope of 1.14, and the other 6 had slopes of $<1$. These results are consistent with previous studies of \cite{2000ApJ...539..300S} and \cite{2021MNRAS.501.4490K} suggesting an extended scattering media along the line of sight. The highest slope of 1.4 is seen by the lowest DM pulsar J1705--1903 out of the 7 MSPs, suggesting a thin scattering screen for lower DM pulsars.
This MSP is unique in our sample, as it is a black widow \citep{2019MNRAS.483.3673M}, and known to be prone
to interactions with material in the near vicinity of the pulsar that
cause depolarisation and absorption. Some of the flux density variations
in this pulsar are thus localized to the source.

\begin{figure}
\centering
\includegraphics[width=0.45\textwidth]{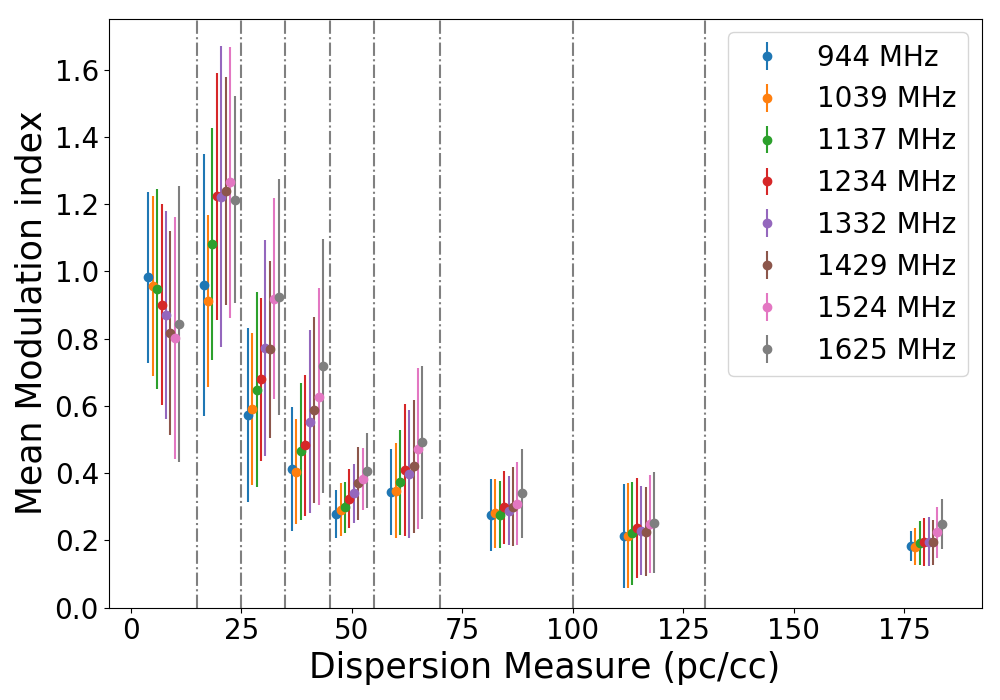}
\caption{ Variation of modulation index with DM of 89 MSPs at 8 different centre observing frequencies. The mean value of the modulation index is calculated for every pulsar in each bin where
the DM boundaries are 0, 15, 25, 35, 45, 55, 70, 100, 130 and 240 pc cm$^{-3}$ shown as dotted vertical lines.
The mean modulation index values for the different frequencies 
are shown with a small offset from the true bin centre to show the error bars that represent the standard deviations more clearly.
}
\label{fig:modind}
\end{figure}

\section{Discussion}
\label{sec:discussion}

An inspection of Table \ref{tab:mod_ind} reveals that 
the average modulation indices vary as a function of DM and in Figure \ref{fig:modind}
we plot the means and standard deviations in 9 DM bins for our 89 MSPs.
At 944 MHz the mean modulation index starts near unity and drops off to about a third of that
value by a DM of 50 pc cm$^{-3}$ before settling at 0.2 at the highest DMs.
At 1625 MHz the peak is shifted to slightly higher DMs, starting at 0.8, peaking at 1.2
by a DM of $\sim$ 25 pc cm$^{-3}$ before dropping to about 0.45 by a DM of 60 pc cm$^{-3}$
and finally reaching a value of near 0.2 at the highest DMs.
This behaviour is similar to that seen by other authors. 
Both
\cite{2000ApJ...539..300S} and
\cite{2018MNRAS.473.4436J} showed that the modulation index of `normal' pulsars decrease with an increase in DM
and are consistent with our results for the MSPs as shown in Figure \ref{fig:modind}. 
Since we are probing the effects of IISM, not the intrinsic pulsar emission, we would expect both `normal' pulsars and MSPs to have consistent modulation index variation with DM. 
The mean modulation index is always lower at the low frequency than the higher frequency equivalent except for the lowest DM bins where the flux density variations are higher. For bins with very large DMs, the mean modulation index is similar for all the observing frequencies. As we go towards higher frequencies, the scattering is weaker hence the modulation index is higher at higher frequencies in comparison to its lower frequency counterpart. \cite{2023MNRAS.tmp..221W} have shown that there exists a negative correlation between modulation index and the assumed 
distance of pulsars,
but as distance is a proxy for DM, this is  consistent with our
findings.

We identify three broad regions in Figure \ref{fig:modind}. The first 
bin is populated by the very low DM pulsars which exhibit a wide range
of modulation indices. The mean modulation indices peak around a DM
of 20 pc cm$^{-3}$ before reducing to more stable
values near 0.2 at high DMs in excess of 100 pc cm$^{-3}$.
For all of the pulsars with a DM greater than 100 pc cm$^{-3}$, the mean modulation index is less than $\sim$0.4 which is consistent with the results from \cite{2023MNRAS.tmp..221W} who have done similar analysis on a sample of both slow and millisecond pulsars.  
Our study has therefore shown that like the slow pulsars, 
MSPs have low flux density modulation indices
at high DMs and are therefore almost certain to have
stable intrinsic radio luminosities (in the absence of intrinsic properties such as eclipses).
Scintillation theory \citep[e.g.][]{1968Natur.218..920S, 1986ApJ...310..737C,2001Ap&SS.278....5R} predicts different scattering regimes
for pulsars that lead to different behaviours. 
The closest pulsars (low DMs)
are close to the regime of weak scintillation, where the difference in the path lengths
travelled by different rays are less than 1 radian in phase and
there are only mild changes in the observed mean fluxes. At 20\,cm
wavelengths such a pulsar is PSR J0437-4715 with a DM = 2.64 pc cm$^{-3}$.
Its time series is shown in Figure \ref{fig:mjdflx}(a) and a histogram of its flux
densities in Figure \ref{fig:histogram}(a). 
In a single observation, the pulsar always exhibits a single continuous mildly modulated scintle spanning the whole spectrum as shown in Figure \ref{fig:psr_scint}(a). Here, the scintillation bandwidth can be much greater than our full observing frequency band and hence, although its flux density varies, its modulation index is quite modest.
Curiously, despite its
very similar DM of 3.14 pc cm$^{-3}$ and distance of $\sim$ 400 pc, 
PSR J1744--1134 has a large modulation index of 1.0(2) at 1429 MHz.
Clearly on 150-400 pc scales particularly at low galactic latitudes the Galaxy can have very different diffractive properties.

Higher DM pulsars like PSR J1909--3744 (DM = 10.39 pc cm$^{-3}$) are in the strong scintillation regime for our observing frequencies. In these cases, the scintillation bandwidth is less than our observed spectral range often causing one or more bright scintles to appear as shown in Figure \ref{fig:psr_scint}(b). 
In this strong scintillation regime, 
the effects are extreme, causing amplified flux intensities but also sometimes nulls. 
This leads to the exponential distribution that is shown in Figure \ref{fig:histogram}(b)
for PSR J1909--3744. The time series plot for this pulsar in Figure \ref{fig:mjdflx}(b) shows seemingly random pulsar flux densities
causing its high observed modulation index of up to 2.0(4) at 1625 MHz. 

As the DM increases further e.g. for PSR J1017--7156 with DM = 94.22 pc cm$^{-3}$, the scintillation bandwidth is reduced further and a multitude of small-bandwidth scintles appear throughout the frequency spectrum (Figure \ref{fig:psr_scint}(c)). The averaging of flux densities in our bands results in more stable fluxes and a lower modulation index. Because of the central
limit theorem the flux density histograms tend towards a Gaussian distribution but are still slightly skewed as seen in Figure \ref{fig:histogram}(c).

For very high DM pulsars like PSR J1747--4036, the scintillation bandwidth gets even smaller and 
small bandwidth scintles become so numerous that they are almost unresolved the flux density becomes almost continuous across the observing bandwidth (see Figure \ref{fig:psr_scint}(d)). Here slow variations in the flux densities arise due to the 
focussing and defocussing of the radio waves through the IISM
via the process of refractive scintillation. The flux density histogram for this pulsar in Figure \ref{fig:histogram}(d) displays a narrower fractional width Gaussian distribution. The slow variations of flux densities can be explored to estimate the refractive scintillation timescales which are described in detail in subsection \ref{sec:stru_func}. The modulation index of PSR J1747--4036 is 0.15(1) at a centre frequency of 1429 MHz which is much smaller than that the
1.3(2) seen for PSR J1909--3744. The value of the structure function at a time lag of
2 weeks is very small and only 0.020(3).
Our results are consistent with \cite{2000ApJ...539..300S} and \cite{2023MNRAS.tmp..221W} who showed that the high DM pulsars 
have low modulation indices and are probably of near constant
intrinsic luminosity. Our high DM pulsar low modulation indices and
structure function values also give us confidence that our flux density calibration methods are
reliable.

\begin{figure*}
\centering
\includegraphics[width=0.85\textwidth]{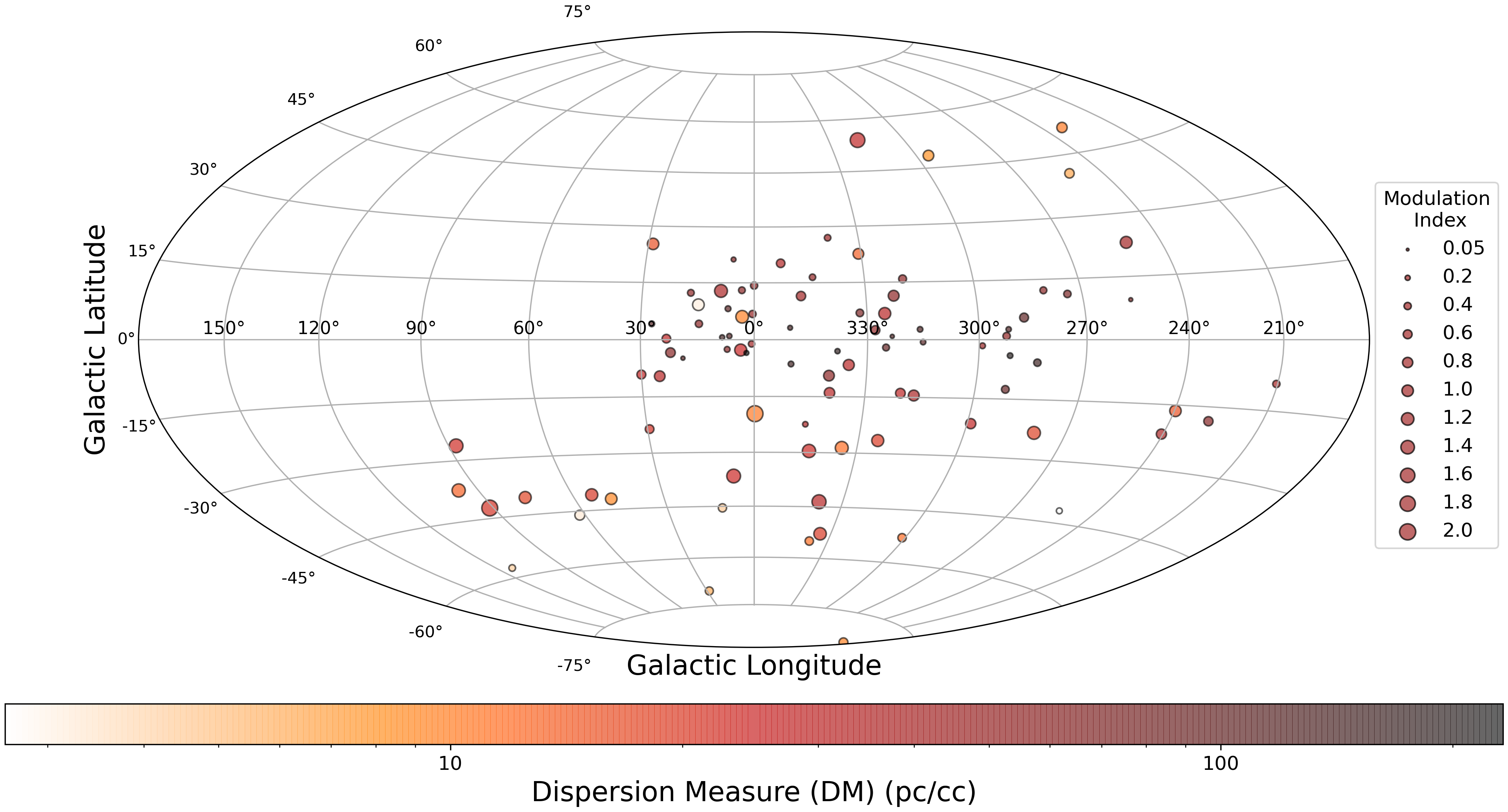}
\caption{ Aitoff-Hammer projection of the 89 MSPs in our sample. 
The area of the dots is proportional to their 1429 MHz modulation index and their DMs are 
represented by different colours as per the legend.
The MSPs between Galactic latitudes of $15^\circ<|b|<45^\circ$ show high variability in their flux densities due to the nature of the scattering screens along their lines of sight.}
\label{fig:hammer}
\end{figure*}

It is interesting to examine how the pulsar modulation indices vary
as a function of Galactic location.
Figure \ref{fig:hammer} shows the Aitoff-Hammer projection of all 89 MSPs in our sample. MSPs along the Galactic plane
usually have low modulation indices because of their higher DMs in comparison to those in the off plane region. This makes sense as there is more clumpy ionised plasma in the Galactic plane containing the Galaxy's spiral arms in comparison to the off plane region where there is much less material  and scattering along the line of sight to the pulsars.
It is very rare that DM $\sin b$, where $b$ is the Galactic latitude, exceeds 25\,pc\, cm$^{-3}$
whereas pulsars in the Galactic plane often have DMs greater than 100\,pc\,cm$^{-3}$ and sometimes over 1000\,pc\,cm$^{-3}$. 
A higher modulation index value is seen between Galactic latitudes $15^\circ<|b|<45^\circ$ 
and thus represents the most highly scintillating part of the Galactic sky at these wavelengths. 
There are a few pulsars that lie in the Galactic plane but still exhibit high variability 
despite having a high DM, presumably due to some combination of their kinematics
and makeup on the ISM along their line of sight. For instance PSR J1103--5403 has a modulation 
index of 0.54(8) even though its DM is 103.91 pc cm$^{-3}$, but PSR J1843--1448
has only a slightly higher DM of 114.54 pc cm$^{-3}$ (both are located at similar angular displacement from the Galactic plane) but
a modulation index of only 0.11(1) at the same frequency. 
PSR J1103--5403 is known to be mode-changing \citep{2023arXiv230402793N}, which is clearly seen when we explore the timing residuals of this pulsar and the residuals can be grouped into two distinct emission `modes'. To explore if the relatively high modulation index of this pulsar is due to the different 
modes, we calculated the flux densities of the observations in each of the two modes and found that the median flux density of the ``upper'' mode was 1.8 times that of the ``lower'' mode (where the timing residuals are higher in the ``upper'' mode than the ``lower''). The modulation indices of
the two modes were 0.55(9) and 0.44(9) for the upper and lower modes respectively, whereas the modulation index of all of the observations was 0.60(8) which demonstrates the effect of emission modes on the results.


\subsection{Implications for PTA observing strategy}\label{sec:obs_strat}

\begin{figure}
\centering
\includegraphics[width=0.45\textwidth]{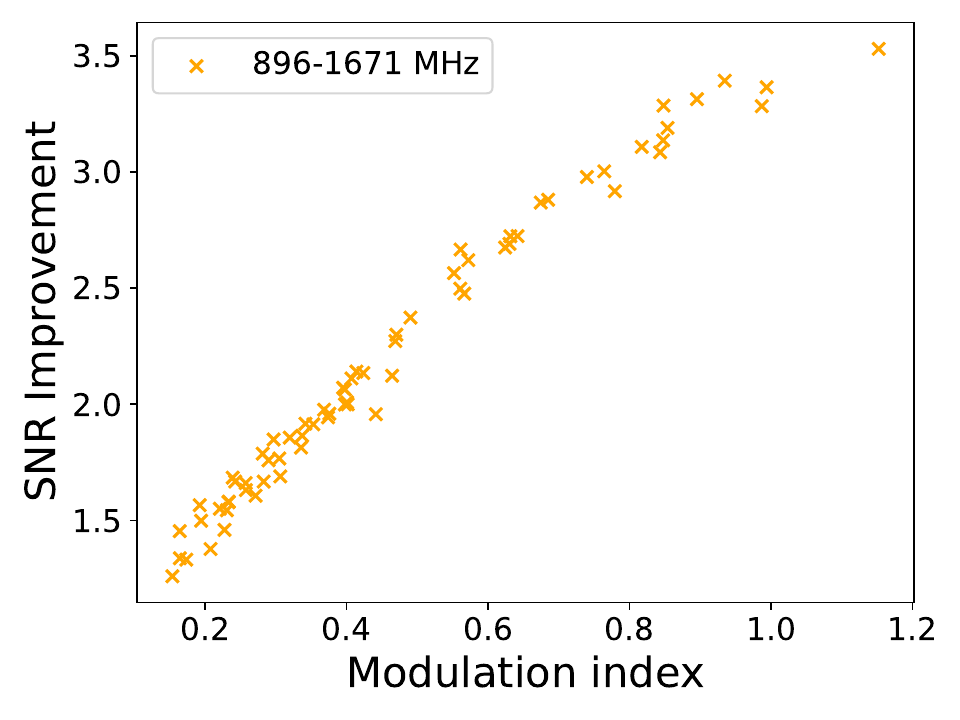}
\caption{Improvement in the SNR of the detected GWB versus flux density 
modulation index of the 
pulsars in our MSP sample with more than 60 observations for the full frequency band.}
\label{fig:obs_strat}
\end{figure}

Our structure functions, modulation indices and
ratios of peak to median fluxes for our 89 MSPs allow us to
explore whether dynamic scheduling, in which we can
dwell longer on a source in a favourable scintillation
state can improve a PTA's timing precision.
The radiometer signal-to-noise's dependence upon flux (linear) and integration time ($\propto t^{1/2}$) (Eqn \ref{eq:flux})
provides the motivation. Pulsars in a high flux state many times
their median are thus far more valuable. 
Unfortunately, many pulsars are jitter-limited and do not
necessarily benefit from increased SNR \citep[e.g.][]{2011MNRAS.418.1258O,  2014MNRAS.443.1463S,  2021MNRAS.502..407P} but nevertheless we can compute what benefits we 
might see for pulsars with very low amounts of jitter. Following \cite{2013CQGra..30v4015S}, the signal to noise ratio $snr$ of detection of gravitational wave background (GWB) in PTAs in ``weak signal limit'' is given by
\begin{equation}
    \label{eqn:gwb}
    snr \propto M c \dfrac{A^{2}}{\sigma^{2}} T^{\beta}
\end{equation}
\noindent where $M$ is the number of pulsars in the PTA, $c$ and $T$ are the cadence and timespan of observations, $A$ and $\beta$ are the amplitude and spectral index of the GWB and $\sigma$ is the error in the times of arrivals (ToAs) of the pulsars. We can estimate the effect of different observation strategies on the detectability of GWB by a PTA using Eqn \ref{eqn:gwb}.
We considered a strategy where any time observers find a flux density in
the top quartile for that pulsar they dwell on it for 4 times as long,
at the expense of three other pulsars that are in a low flux density state.
For each of the 89 pulsars in the MPTA we therefore computed the ratio of the sum of
4 times the signal-to-noise ratios squared of the top quartile of observations,
and compared them to that of all the observations across the entire MeerKAT observing
band. This number, between 1-4,
gives the improvement in the gravitational wave detection signal-to-noise ratio and is plotted in Figure \ref{fig:obs_strat}
as a function of the modulation index. 
The choice of the top quartile is somewhat arbitrary, but if the objective is to detect
low-frequency gravitational waves it is probably necessary to observe at
least 6 times per year, and this is about 1/4 of the current cadence being
achieved for the MPTA.
From our Figure we can see that even pulsars with a modest modulation index of 0.2 achieve
a 50\% improvement in detection signal-to-noise and those near unity are closer to a factor of 3.25.

MeerKAT will eventually become part of the 196-dish SKA-Mid.
One could imagine a 16 element sub-array quickly performing
a census of the flux state of the next few pulsars in the schedule and
then joining the rest of the antennas to monitor the
pulsars in favourable scintillation states \citep{2021MNRAS.502..407P}. 
According to our current understanding, the GWB is only present
at very low (nHz) frequencies, hence a shorter cadence won't affect the Hellings and Downs cross-correlation curve presence required for the detection of GWs.
Another potential penalty that might arise from ``scint-hunting'' is that a small
uncertainty in the arrival time in one sub-band is
only of value if the DM can be accurately determined,
and this is sometimes compromised when there is insufficient flux
across the band to accurately determine the DM. In such cases
the uncertainty in the DM dominates the arrival times.

\subsection{Implications of the IISM for FRB discovery rates and FRB luminosities} \label{sec:frb_lum}

FRBs are transient events in the sky that are of unknown astronomical origin which are highly energetic, on time scales of $\mu$s to tens of milliseconds
\citep[e.g.][]{2007Sci...318..777L, 2022Sci...378.3043B}.
Because of their immense distances, FRBs are expected to be point sources and like pulsars
the IISM is expected to amplify the flux densities of FRBs, and might affect their observed population
at different galactic latitudes \citep{2015MNRAS.451.3278M, 2017MNRAS.468.2726K}. 
We expect scintillation to more frequently amplify FRBs at high Galactic latitudes than near the plane, as shown in simulations by \citet{2014ApJ...789L..26P}.
Hence, the inferred luminosities of especially one-off FRBs at high latitudes might be systematically overestimated in FRB surveys.

We studied this using Monte Carlo simulations of synthetic populations of cosmologically distributed FRBs, assuming a flux-limited detection threshold, including scintillation effects due to the IISM. Each FRB was placed in a volume up to a maximum distance well beyond our detection threshold. The FRB luminosity function is known to have a wide energy distribution as described in \cite[e.g.][]{2022MNRAS.509.4775J, 2023ApJ...944..105S}. 
For simplicity we 
considered a log-normal distribution as the extent of the luminosity
function is more important than its shape and assigned each simulated FRB 
an intrinsic luminosity and calculated the flux density. 
This flux density was then multiplied by a magnification factor which was drawn from an exponential distribution (similar to that of PSR J1909--3744 - see Figure \ref{fig:histogram} (b)), to represent FRBs affected by strong scintillation where it is well approximated by an exponential.  
If the scintillated flux was greater than our detection threshold we recorded it as a detection,
and the luminosity, distance and flux recorded.

We found that in the exponential-distribution scintillation case, the average distance up to which we can detect FRBs is $\sim$ 2.35 times more 
than in the no-scintillation case. Therefore surveys in directions where scintillation is as high as seen in PSR J1909--3744 could have FRB discovery rates increased by $\sim$32\%. 
We also found that on average the FRBs in the exponential-scintillation case have their mean luminosities overestimated by a factor of about 2.4. 
But is this borne out by current FRB surveys?
Interestingly, in by far the most prolific FRB survey undertaken by CHIME, \cite{2021ApJ...923....2J} found no evidence for any dependence between FRB detection and Galactic latitude by performing statistical tests on CHIME FRBs taking into account many selection effects such as telescope sensitivity, dwell time with declination etc. They found that in the CHIME observing frequency band (400--800 MHz), strong scintillation effects, of the type we see in this sample, are not a factor.
Our simulations predicted that FRB surveys will be $\sim$30\% more prolific
if they survey regions of the Galaxy prone to high degrees of 
scintillation or at a higher observing frequency and it will be interesting to see if this is borne
out by future surveys at frequencies similar to that of the MPTA data such as that of TRAPUM \citep{2016mks..confE...9S} and MeerTRAP \citep{2016mks..confE..10S}.

\section{Conclusions}
\label{sec:conclusion}
We presented the flux density variability of 89 MSPs 
based upon regular observations of the MeerKAT Pulsar Timing Array. 
Despite the lack of a pulsed cal, we were able to show that our
derived flux densities are reliable and suitable for modulation index
and structure function analyses. Our study demonstrates
that MSPs on the whole have simple power law spectra over the range
856--1712 MHz, with only a very modest tendency for
any curvature. We demonstrated that the MSPs all vary in flux density,
but this is a strong function of their DMs and indicates that
MSPs have intrinsically stable radio luminosities.
Our results will help in planning future timing array and
Shapiro delay campaigns with both MeerKAT and other existing 
and future 20\,cm
band radio telescopes such as the SKA and the Deep Synoptic
Array 2000. Pulsars with high DMs and long refractive scintillation
timescales can have their flux densities reliably predicted
from epoch to epoch.
Our variability metrics can also be used to optimise timing array observation strategies.
Finally, our simulations predict that 20\,cm FRB surveys should
have more success at finding distant FRBs at the mid-galactic latitudes where the modulation indices of MSPs peak.

\begin{table*}
\centering
\caption{Measured flux density structure functions of the 89 MSPs in our sample at a central frequency of 1429 MHz at a time lag of 2 weeks, 1 month, 1 year and 3 years. }
\label{tab:stru_func}

\begin{tabular}{rrrrrr}
\hline
\multicolumn{1}{c} {NAME}   & \multicolumn{1}{c}{DM}  &  \multicolumn{1}{c}{$\rm D(\rm \tau)$}  & \multicolumn{1}{c}{$ \rm D(\rm \tau)$}  & \multicolumn{1}{c}{$ \rm D(\rm \tau)$}  &  \multicolumn{1}{c}{$\rm  D(\rm \tau)$}    \\
& \multicolumn{1}{c}{(pc cm$^{-3}$)} & \multicolumn{1}{c}{(2 weeks)} & \multicolumn{1}{c}{(1 month)} & \multicolumn{1}{c}{(1 year)} & \multicolumn{1}{c}{(3 yrs)} \\
\hline
J0030+0451 & 4.33   & 0.258(7)  & 0.326(3)  & 0.292(1)   & 0.318(2)   \\
J0101$-$6422 & 11.92  & 0.765(4)  & 0.486(1)  & 0.5703(9)  & 0.554(1)   \\
J0125$-$2327 & 9.6    & 0.63(2)   & 0.80(1)   & 0.941(4)   & 0.875(4)   \\
J0437$-$4715 & 2.64   & 0.2(7)    & 0.2(3)    & 0.2(2)     & 0.2(2)     \\
J0610$-$2100 & 60.69  & 0.434(5)  & 0.690(3)  & 0.611(1)   & 0.798(1)   \\
J0613$-$0200 & 38.78  & 0.175(8)  & 0.381(5)  & 0.306(2)   & 0.283(2)   \\
J0614$-$3329 & 37.05  & 1.179(8)  & 1.329(3)  & 1.469(2)   & 1.253(1)   \\
J0636$-$3044 & 15.46  & 1.91(4)   & 2.86(2)   & 2.911(9)   & 2.511(9)   \\
J0711$-$6830 & 18.41  & 2.32(6)   & 6.44(4)   & 5.26(2)    & 5.02(2)    \\
J0900$-$3144 & 75.69  & 0.009(4)  & 0.011(2)  & 0.027(1)   & 0.022(1)   \\
J0931$-$1902 & 41.49  & 1.628(8)  & 0.929(3)  & 1.304(2)   & 2.721(3)   \\
J0955$-$6150 & 160.9  & 0.138(3)  & 0.129(1)  & 0.2235(8)  & 0.2162(8)  \\
J1012$-$4235 & 71.65  & 0.2260(6) & 0.2296(3) & 0.2452(2)  & 0.2404(2)  \\
J1017$-$7156 & 94.22  & 0.270(3)  & 0.299(1)  & 0.3553(6)  & 0.3962(7)  \\
J1022+1001 & 10.25  & 1.66(4)   & 1.70(2)   & 1.578(6)   & 1.661(7)   \\
J1024$-$0719 & 6.49   & 1.16(2)   & 0.913(7)  & 1.116(3)   & 0.947(3)   \\
J1036$-$8317 & 27.09  & 0.749(4)  & 1.182(2)  & 1.576(1)   & 1.264(1)   \\
J1045$-$4509 & 58.11  & 0.18(1)   & 0.191(5)  & 0.233(2)   & 0.226(2)   \\
J1101$-$6424 & 207.36 & 0.0257(7) & 0.0452(4) & 0.0603(3)  & 0.0760(4)  \\
J1103$-$5403 & 103.91 & 0.525(2)  & 0.554(1)  & 0.5762(5)  & 0.6016(5)  \\
J1125$-$5825 & 124.82 & 0.061(3)  & 0.060(1)  & 0.0896(9)  & 0.138(1)   \\
J1125$-$6014 & 52.93  & 0.264(9)  & 0.249(4)  & 0.242(2)   & 0.304(2)   \\
J1216$-$6410 & 47.39  & 0.133(6)  & 0.136(3)  & 0.131(1)   & 0.141(1)   \\
J1231$-$1411 & 8.09   & 1.338(6)  & 1.600(6)  & 2.376(6)   & 1.89(3)    \\
J1327$-$0755 & 27.91  & 2.947(4)  & 2.290(2)  & 3.024(1)   & 2.745(1)   \\
J1421$-$4409 & 54.64  & 0.33(1)   & 0.400(5)  & 0.397(2)   & 0.347(2)   \\
J1431$-$5740 & 131.38 & 0.0213(4) & 0.0739(3) & 0.1734(2)  & 0.1195(2)  \\
J1435$-$6100 & 113.78 & 0.0707(4) & 0.0684(2) & 0.08210(9) & 0.08636(9) \\
J1446$-$4701 & 55.83  & 0.929(3)  & 0.936(2)  & 0.8870(7)  & 0.9578(7)  \\
J1455$-$3330 & 13.57  & 0.749(6)  & 2.247(4)  & 2.404(2)   & 2.475(2)   \\
J1513$-$2550 & 46.88  & 0.074(2)  & 0.101(1)  & 0.1288(8)  & 0.144(1)   \\
J1514$-$4946 & 31.01  & 0.418(2)  & 0.952(1)  & 1.3388(7)  & 1.394(1)   \\
J1525$-$5545 & 126.97 & 0.0111(3) & 0.0106(2) & 0.0255(1)  & 0.0515(2)  \\
J1543$-$5149 & 50.98  & 0.46(1)   & 0.638(5)  & 0.795(2)   & 0.549(2)   \\
J1545$-$4550 & 68.39  & 0.344(8)  & 0.283(3)  & 0.267(1)   & 0.305(2)   \\
J1547$-$5709 & 95.72  & 0.170(2)  & 0.244(1)  & 0.2127(8)  & 0.2116(8)  \\
J1600$-$3053 & 52.33  & 0.203(7)  & 0.185(3)  & 0.173(1)   & 0.164(1)   \\
J1603$-$7202 & 38.05  & 1.13(3)   & 1.43(1)   & 1.139(5)   & 1.283(6)   \\
J1614$-$2230 & 34.49  & 0.278(9)  & 0.402(4)  & 0.424(2)   & 0.435(2)   \\
J1628$-$3205 & 42.14  & 0.42(1)   & 1.26(1)   & 0.503(5)   & 0.631(4)   \\
J1629$-$6902 & 29.49  & 0.682(6)  & 0.933(3)  & 0.821(1)   & 0.781(1)   \\
J1643$-$1224 & 62.4   & 0.008(5)  & 0.013(3)  & 0.021(2)   & 0.055(2)   \\
J1652$-$4838 & 188.16 & 0.033(2)  & 0.047(1)  & 0.0497(7)  & 0.0682(7)  \\
J1653$-$2054 & 56.52  & 0.217(4)  & 0.308(2)  & 0.328(1)   & 0.273(1)   \\
J1658$-$5324 & 30.83  & 2.73(2)   & 2.000(7)  & 2.273(3)   & 1.963(4)   \\
J1705$-$1903 & 57.51  & 0.084(2)  & 0.205(1)  & 0.2595(6)  & 0.2819(8)  \\
J1708$-$3506 & 146.77 & 0.013(3)  & 0.012(2)  & 0.013(1)   & 0.033(2)   \\
J1713+0747 & 15.99  & 1.07(6)   & 2.55(4)   & 2.28(2)    & 2.66(2)    \\
J1719$-$1438 & 36.78  & 1.747(5)  & 1.237(2)  & 1.0497(7)  & 1.1058(8)  \\
J1721$-$2457 & 48.23  & 0.348(9)  & 0.331(4)  & 0.394(2)   & 0.364(2)   \\
J1730$-$2304 & 9.63   & 0.94(5)   & 1.75(3)   & 1.60(1)    & 1.40(1)    \\
J1731$-$1847 & 106.47 & 0.074(4)  & 0.112(3)  & 0.107(2)   & 0.136(3)   \\
J1732$-$5049 & 56.82  & 1.79(3)   & 1.97(1)   & 1.727(5)   & 1.272(5)   \\
J1737$-$0811 & 55.3   & 0.071(4)  & 0.121(2)  & 0.129(1)   & 0.152(1)   \\
J1744$-$1134 & 3.14   & 2.01(4)   & 2.01(2)   & 2.140(8)   & 1.908(8)   \\
J1747$-$4036 & 152.94 & 0.020(3)  & 0.024(2)  & 0.047(1)   & 0.049(1)   \\
J1751$-$2857 & 42.79  & 0.138(2)  & 0.1521(9) & 0.1599(5)  & 0.1365(5)  \\
J1756$-$2251 & 121.23 & 0.0095(9) & 0.0141(5) & 0.0519(4)  & 0.1154(5)  \\
J1757$-$5322 & 30.8   & 0.389(9)  & 0.461(4)  & 0.502(2)   & 0.485(2)   \\
J1801$-$1417 & 57.25  & 0.199(9)  & 0.248(4)  & 0.319(2)   & 0.284(2)   \\
J1802$-$2124 & 149.59 & 0.0119(6) & 0.0253(4) & 0.0617(3)  & 0.0724(3)  \\
J1804$-$2717 & 24.67  & 0.95(3)   & 3.47(3)   & 3.08(1)    & 2.05(2)    \\
\hline
\end{tabular}
\end{table*}

\begin{table*}
\centering

\begin{tabular}{rrrrrr}
\hline
\multicolumn{1}{c} {NAME}   & \multicolumn{1}{c}{DM}  &  \multicolumn{1}{c}{$\rm D(\rm \tau)$}  & \multicolumn{1}{c}{$ \rm D(\rm \tau)$}  & \multicolumn{1}{c}{$ \rm D(\rm \tau)$}  &  \multicolumn{1}{c}{$\rm  D(\rm \tau)$}    \\
 &\multicolumn{1}{c}{(pc cm$^{-3}$)} & \multicolumn{1}{c}{(2 weeks)} & \multicolumn{1}{c}{(1 month)} & \multicolumn{1}{c}{(1 year)} & \multicolumn{1}{c}{(3 yrs)} \\
 \hline 
J1804$-$2858 & 232.52 & 0.036(5)  & 0.025(2)  & 0.029(2)   & 0.026(2)  \\
J1811$-$2405 & 60.62  & 0.051(4)  & 0.078(2)  & 0.092(1)   & 0.102(1)   \\
J1825$-$0319 & 119.56 & 0.0389(6) & 0.0484(4) & 0.0614(3)  & 0.0801(3)  \\
J1832$-$0836 & 28.19  & 1.12(2)   & 0.825(7)  & 1.27(4)    & 1.7(1)     \\
J1843$-$1113 & 59.96  & 0.467(4)  & 0.386(1)  & 0.4160(6)  & 0.4932(7)  \\
J1843$-$1448 & 114.54 & 0.013(1)  & 0.0161(9) & 0.0223(7)  & 0.0277(9)  \\
J1902$-$5105 & 36.25  & 0.060(2)  & 0.077(1)  & 0.0698(6)  & 0.0736(6)  \\
J1903$-$7051 & 19.66  & 2.18(1)   & 2.165(4)  & 2.410(2)   & 2.416(2)   \\
J1909$-$3744 & 10.39  & 4.70(1)   & 3.259(3)  & 3.279(1)   & 3.189(1)   \\
J1911$-$1114 & 30.97  & 1.05(2)   & 1.306(8)  & 0.996(3)   & 0.683(4)   \\
J1918$-$0642 & 26.59  & 0.30(1)   & 0.301(5)  & 0.307(2)   & 0.282(2)   \\
J1933$-$6211 & 11.51  & 3.81(2)   & 3.909(8)  & 4.324(3)   & 3.604(3)   \\
J1946$-$5403 & 23.73  & 3.566(4)  & 5.948(2)  & 5.2545(9)  & 4.7866(9)  \\
J2010$-$1323 & 22.16  & 0.422(2)  & 0.3437(9) & 0.4034(4)  & 0.4727(4)  \\
J2039$-$3616 & 23.96  & 2.427(8)  & 3.112(4)  & 3.222(2)   & 2.995(2)   \\
J2124$-$3358 & 4.6    & 0.80(6)   & 0.72(2)   & 0.71(1)    & 0.58(1)    \\
J2129$-$5721 & 31.85  & 1.79(1)   & 3.028(6)  & 3.224(3)   & 2.991(3)   \\
J2145$-$0750 & 9.0    & 4.7(1)    & 3.03(5)   & 2.99(2)    & 2.81(2)    \\
J2150$-$0326 & 20.67  & 2.340(5)  & 2.298(2)  & 1.9272(8)  & 2.772(1)   \\
J2222$-$0137 & 3.28   & 1.394(7)  & 0.982(2)  & 1.120(1)   & 1.154(1)   \\
J2229+2643 & 22.73  & 6.28(2)   & 8.04(1)   & 8.124(4)   & 7.506(5)   \\
J2234+0944 & 17.83  & 1.78(3)   & 3.01(1)   & 2.754(6)   & 2.447(6)   \\
J2236$-$5527 & 20.09  & 2.111(5)  & 2.529(3)  & 2.352(1)   & 3.404(2)   \\
J2241$-$5236 & 11.41  & 0.71(1)   & 0.786(5)  & 0.692(2)   & 0.772(2)   \\
J2317+1439 & 21.9   & 3.97(1)   & 3.547(5)  & 3.513(2)   & 3.848(2)   \\
J2322+2057 & 13.38  & 3.749(7)  & 2.302(2)  & 2.186(1)   & 2.256(1)   \\
J2322$-$2650 & 6.15   & 0.468(2)  & 0.4854(9) & 0.4967(5)  & 0.5032(5) \\
\hline
\end{tabular}
\end{table*}

\section*{Acknowledgements}
The MeerKAT telescope
is operated by the South African Radio Astronomy Observatory,
which is a facility of the National Research Foundation, an agency
of the Department of Science and Innovation. PG acknowledges the support of an SUT post graduate stipend. RMS, MB
and DJR acknowledge support through the Australian Research Council (ARC) centre of Excellence
grant CE17010004 (OzGrav).
RMS acknowledges support through
ARC Future Fellowship FT190100155. 
This work used
the OzSTAR national facility at Swinburne University of Technology
and the pulsar portal maintained by ADACS at URL: https://pulsars.org.au 
OzSTAR and the pulsar portal are funded by Swinburne University of Technology and the
National Collaborative Research Infrastructure Strategy (NCRIS).

\section*{Data Availability}
The data set used here will be made available along with the manuscript under the DOI \url{http://dx.doi.org/10.26185/6487ea315602c}.
 For each MSP, we have included PSRFITs files of pulsar observations which are fully time and polarization averaged and the frequency channels are averaged into 8 sub-bands.
 We have provided the mean, median flux densities, standard deviations and modulation indices at 8 different frequency bands for all 89 MSPs in a csv file format. Spectral indices and pulsar parameters such as period, global coordinates and DM are also included in the data set.
 The data set is also available in the MeerTime pulsar portal \url{pulsars.org.au} which contains all of the MSP observations so far, their raw and cleaned profiles, timing residuals and their dynamic spectra. The data is publicly available after 18 months of embargo.  


\bibliographystyle{mnras}
\bibliography{example} 

\begin{thebibliography}{}
\makeatletter
\relax
\def\mn@urlcharsother{\let\do\@makeother \do\$\do\&\do\#\do\^\do\_\do\%\do\~}
\def\mn@doi{\begingroup\mn@urlcharsother \@ifnextchar [ {\mn@doi@} {\mn@doi@[]}}
\def\mn@doi@[#1]#2{\def\@tempa{#1}\ifx\@tempa\@empty \href {http://dx.doi.org/#2} {doi:#2}\else \href {http://dx.doi.org/#2} {#1}\fi \endgroup}
\def\mn@eprint#1#2{\mn@eprint@#1:#2::\@nil}
\def\mn@eprint@arXiv#1{\href {http://arxiv.org/abs/#1} {{\tt arXiv:#1}}}
\def\mn@eprint@dblp#1{\href {http://dblp.uni-trier.de/rec/bibtex/#1.xml} {dblp:#1}}
\def\mn@eprint@#1:#2:#3:#4\@nil{\def\@tempa {#1}\def\@tempb {#2}\def\@tempc {#3}\ifx \@tempc \@empty \let \@tempc \@tempb \let \@tempb \@tempa \fi \ifx \@tempb \@empty \def\@tempb {arXiv}\fi \@ifundefined {mn@eprint@\@tempb}{\@tempb:\@tempc}{\expandafter \expandafter \csname mn@eprint@\@tempb\endcsname \expandafter{\@tempc}}}

\bibitem[\protect\citeauthoryear{{Agazie} et~al.,}{{Agazie} et~al.}{2023}]{2023ApJ...951L...9A}
{Agazie} G.,  et~al., 2023, \mn@doi [\apjl] {10.3847/2041-8213/acda9a}, \href {https://ui.adsabs.harvard.edu/abs/2023ApJ...951L...9A} {951, L9}

\bibitem[\protect\citeauthoryear{{Antoniadis} et~al.,}{{Antoniadis} et~al.}{2023}]{2023arXiv230616224A}
{Antoniadis} J.,  et~al., 2023, \mn@doi [arXiv e-prints] {10.48550/arXiv.2306.16224}, \href {https://ui.adsabs.harvard.edu/abs/2023arXiv230616224A} {p. arXiv:2306.16224}

\bibitem[\protect\citeauthoryear{{Armstrong}, {Rickett}  \& {Spangler}}{{Armstrong} et~al.}{1995}]{1995ApJ...443..209A}
{Armstrong} J.~W.,  {Rickett} B.~J.,   {Spangler} S.~R.,  1995, \mn@doi [\apj] {10.1086/175515}, \href {https://ui.adsabs.harvard.edu/abs/1995ApJ...443..209A} {443, 209}

\bibitem[\protect\citeauthoryear{{Askew}, {Reardon}  \& {Shannon}}{{Askew} et~al.}{2023}]{2023MNRAS.519.5086A}
{Askew} J.,  {Reardon} D.~J.,   {Shannon} R.~M.,  2023, \mn@doi [\mnras] {10.1093/mnras/stac3095}, \href {https://ui.adsabs.harvard.edu/abs/2023MNRAS.519.5086A} {519, 5086}

\bibitem[\protect\citeauthoryear{{Backer}}{{Backer}}{1970a}]{1970Natur.228...42B}
{Backer} D.~C.,  1970a, \mn@doi [\nat] {10.1038/228042a0}, \href {https://ui.adsabs.harvard.edu/abs/1970Natur.228...42B} {228, 42}

\bibitem[\protect\citeauthoryear{{Backer}}{{Backer}}{1970b}]{1970Natur.228.1297B}
{Backer} D.~C.,  1970b, \mn@doi [\nat] {10.1038/2281297a0}, \href {https://ui.adsabs.harvard.edu/abs/1970Natur.228.1297B} {228, 1297}

\bibitem[\protect\citeauthoryear{{Bailes}}{{Bailes}}{2010}]{2010IAUS..261..212B}
{Bailes} M.,  2010, in Relativity in Fundamental Astronomy: Dynamics, Reference Frames, and Data Analysis. pp 212--217, \mn@doi{10.1017/S1743921309990421}

\bibitem[\protect\citeauthoryear{{Bailes}}{{Bailes}}{2022}]{2022Sci...378.3043B}
{Bailes} M.,  2022, \mn@doi [Science] {10.1126/science.abj3043}, \href {https://ui.adsabs.harvard.edu/abs/2022Sci...378.3043B} {378, abj3043}

\bibitem[\protect\citeauthoryear{{Bailes} et~al.,}{{Bailes} et~al.}{2020}]{2020PASA...37...28B}
{Bailes} M.,  et~al., 2020, \mn@doi [\pasa] {10.1017/pasa.2020.19}, \href {https://ui.adsabs.harvard.edu/abs/2020PASA...37...28B} {37, e028}

\bibitem[\protect\citeauthoryear{{Boyles} et~al.,}{{Boyles} et~al.}{2013}]{2013ApJ...763...80B}
{Boyles} J.,  et~al., 2013, \mn@doi [\apj] {10.1088/0004-637X/763/2/80}, \href {https://ui.adsabs.harvard.edu/abs/2013ApJ...763...80B} {763, 80}

\bibitem[\protect\citeauthoryear{{Cole}, {Hesse}  \& {Page}}{{Cole} et~al.}{1970}]{1970Natur.225..712C}
{Cole} T.~W.,  {Hesse} H.~K.,   {Page} C.~G.,  1970, \mn@doi [\nat] {10.1038/225712a0}, \href {https://ui.adsabs.harvard.edu/abs/1970Natur.225..712C} {225, 712}

\bibitem[\protect\citeauthoryear{{Cordes}}{{Cordes}}{1986}]{1986ApJ...311..183C}
{Cordes} J.~M.,  1986, \mn@doi [\apj] {10.1086/164764}, \href {https://ui.adsabs.harvard.edu/abs/1986ApJ...311..183C} {311, 183}

\bibitem[\protect\citeauthoryear{{Cordes} \& {Downs}}{{Cordes} \& {Downs}}{1985}]{1985ApJS...59..343C}
{Cordes} J.~M.,  {Downs} G.~S.,  1985, \mn@doi [\apjs] {10.1086/191076}, \href {https://ui.adsabs.harvard.edu/abs/1985ApJS...59..343C} {59, 343}

\bibitem[\protect\citeauthoryear{{Cordes}, {Pidwerbetsky}  \& {Lovelace}}{{Cordes} et~al.}{1986}]{1986ApJ...310..737C}
{Cordes} J.~M.,  {Pidwerbetsky} A.,   {Lovelace} R.~V.~E.,  1986, \mn@doi [\apj] {10.1086/164728}, \href {https://ui.adsabs.harvard.edu/abs/1986ApJ...310..737C} {310, 737}

\bibitem[\protect\citeauthoryear{{Dai} et~al.,}{{Dai} et~al.}{2015}]{2015MNRAS.449.3223D}
{Dai} S.,  et~al., 2015, \mn@doi [\mnras] {10.1093/mnras/stv508}, \href {https://ui.adsabs.harvard.edu/abs/2015MNRAS.449.3223D} {449, 3223}

\bibitem[\protect\citeauthoryear{{Detweiler}}{{Detweiler}}{1979}]{1979ApJ...234.1100D}
{Detweiler} S.,  1979, \mn@doi [\apj] {10.1086/157593}, \href {https://ui.adsabs.harvard.edu/abs/1979ApJ...234.1100D} {234, 1100}

\bibitem[\protect\citeauthoryear{{Geyer} et~al.,}{{Geyer} et~al.}{2021}]{2021MNRAS.505.4468G}
{Geyer} M.,  et~al., 2021, \mn@doi [\mnras] {10.1093/mnras/stab1501}, \href {https://ui.adsabs.harvard.edu/abs/2021MNRAS.505.4468G} {505, 4468}

\bibitem[\protect\citeauthoryear{{Haslam}, {Salter}, {Stoffel}  \& {Wilson}}{{Haslam} et~al.}{1982}]{1982A&AS...47....1H}
{Haslam} C.~G.~T.,  {Salter} C.~J.,  {Stoffel} H.,   {Wilson} W.~E.,  1982, \aaps, \href {https://ui.adsabs.harvard.edu/abs/1982A&AS...47....1H} {47, 1}

\bibitem[\protect\citeauthoryear{{Haverkorn}, {Gaensler}, {McClure-Griffiths}, {Dickey}  \& {Green}}{{Haverkorn} et~al.}{2004}]{2004ApJ...609..776H}
{Haverkorn} M.,  {Gaensler} B.~M.,  {McClure-Griffiths} N.~M.,  {Dickey} J.~M.,   {Green} A.~J.,  2004, \mn@doi [\apj] {10.1086/421341}, \href {https://ui.adsabs.harvard.edu/abs/2004ApJ...609..776H} {609, 776}

\bibitem[\protect\citeauthoryear{{Hellings} \& {Downs}}{{Hellings} \& {Downs}}{1983}]{1983ApJ...265L..39H}
{Hellings} R.~W.,  {Downs} G.~S.,  1983, \mn@doi [\apjl] {10.1086/183954}, \href {https://ui.adsabs.harvard.edu/abs/1983ApJ...265L..39H} {265, L39}

\bibitem[\protect\citeauthoryear{{Hewish}, {Bell}, {Pilkington}, {Scott}  \& {Collins}}{{Hewish} et~al.}{1968}]{1968Natur.217..709H}
{Hewish} A.,  {Bell} S.~J.,  {Pilkington} J.~D.~H.,  {Scott} P.~F.,   {Collins} R.~A.,  1968, \mn@doi [\nat] {10.1038/217709a0}, \href {https://ui.adsabs.harvard.edu/abs/1968Natur.217..709H} {217, 709}

\bibitem[\protect\citeauthoryear{{James}, {Prochaska}, {Macquart}, {North-Hickey}, {Bannister}  \& {Dunning}}{{James} et~al.}{2022}]{2022MNRAS.509.4775J}
{James} C.~W.,  {Prochaska} J.~X.,  {Macquart} J.~P.,  {North-Hickey} F.~O.,  {Bannister} K.~W.,   {Dunning} A.,  2022, \mn@doi [\mnras] {10.1093/mnras/stab3051}, \href {https://ui.adsabs.harvard.edu/abs/2022MNRAS.509.4775J} {509, 4775}

\bibitem[\protect\citeauthoryear{{Jankowski}, {van Straten}, {Keane}, {Bailes}, {Barr}, {Johnston}  \& {Kerr}}{{Jankowski} et~al.}{2018}]{2018MNRAS.473.4436J}
{Jankowski} F.,  {van Straten} W.,  {Keane} E.~F.,  {Bailes} M.,  {Barr} E.~D.,  {Johnston} S.,   {Kerr} M.,  2018, \mn@doi [\mnras] {10.1093/mnras/stx2476}, \href {https://ui.adsabs.harvard.edu/abs/2018MNRAS.473.4436J} {473, 4436}

\bibitem[\protect\citeauthoryear{{Johnston}, {Manchester}, {Lyne}, {Bailes}, {Kaspi}, {Qiao}  \& {D'Amico}}{{Johnston} et~al.}{1992}]{1992ApJ...387L..37J}
{Johnston} S.,  {Manchester} R.~N.,  {Lyne} A.~G.,  {Bailes} M.,  {Kaspi} V.~M.,  {Qiao} G.,   {D'Amico} N.,  1992, \mn@doi [\apjl] {10.1086/186300}, \href {https://ui.adsabs.harvard.edu/abs/1992ApJ...387L..37J} {387, L37}

\bibitem[\protect\citeauthoryear{{Josephy} et~al.,}{{Josephy} et~al.}{2021}]{2021ApJ...923....2J}
{Josephy} A.,  et~al., 2021, \mn@doi [\apj] {10.3847/1538-4357/ac33ad}, \href {https://ui.adsabs.harvard.edu/abs/2021ApJ...923....2J} {923, 2}

\bibitem[\protect\citeauthoryear{{Keane} et~al.,}{{Keane} et~al.}{2015}]{2015aska.confE..40K}
{Keane} E.,  et~al., 2015, in Advancing Astrophysics with the Square Kilometre Array (AASKA14). p.~40 (\mn@eprint {arXiv} {1501.00056}), \mn@doi{10.22323/1.215.0040}

\bibitem[\protect\citeauthoryear{{Kramer}, {Xilouris}, {Lorimer}, {Doroshenko}, {Jessner}, {Wielebinski}, {Wolszczan}  \& {Camilo}}{{Kramer} et~al.}{1998}]{1998ApJ...501..270K}
{Kramer} M.,  {Xilouris} K.~M.,  {Lorimer} D.~R.,  {Doroshenko} O.,  {Jessner} A.,  {Wielebinski} R.,  {Wolszczan} A.,   {Camilo} F.,  1998, \mn@doi [\apj] {10.1086/305790}, \href {https://ui.adsabs.harvard.edu/abs/1998ApJ...501..270K} {501, 270}

\bibitem[\protect\citeauthoryear{{Kramer}, {Lange}, {Lorimer}, {Backer}, {Xilouris}, {Jessner}  \& {Wielebinski}}{{Kramer} et~al.}{1999}]{1999ApJ...526..957K}
{Kramer} M.,  {Lange} C.,  {Lorimer} D.~R.,  {Backer} D.~C.,  {Xilouris} K.~M.,  {Jessner} A.,   {Wielebinski} R.,  1999, \mn@doi [\apj] {10.1086/308042}, \href {https://ui.adsabs.harvard.edu/abs/1999ApJ...526..957K} {526, 957}

\bibitem[\protect\citeauthoryear{{Kramer}, {Lyne}, {O'Brien}, {Jordan}  \& {Lorimer}}{{Kramer} et~al.}{2006}]{2006Sci...312..549K}
{Kramer} M.,  {Lyne} A.~G.,  {O'Brien} J.~T.,  {Jordan} C.~A.,   {Lorimer} D.~R.,  2006, \mn@doi [Science] {10.1126/science.1124060}, \href {https://ui.adsabs.harvard.edu/abs/2006Sci...312..549K} {312, 549}

\bibitem[\protect\citeauthoryear{{Kramer} et~al.,}{{Kramer} et~al.}{2021a}]{2021PhRvX..11d1050K}
{Kramer} M.,  et~al., 2021a, \mn@doi [Physical Review X] {10.1103/PhysRevX.11.041050}, \href {https://ui.adsabs.harvard.edu/abs/2021PhRvX..11d1050K} {11, 041050}

\bibitem[\protect\citeauthoryear{{Kramer} et~al.,}{{Kramer} et~al.}{2021b}]{2021MNRAS.504.2094K}
{Kramer} M.,  et~al., 2021b, \mn@doi [\mnras] {10.1093/mnras/stab375}, \href {https://ui.adsabs.harvard.edu/abs/2021MNRAS.504.2094K} {504, 2094}

\bibitem[\protect\citeauthoryear{{Kumamoto} et~al.,}{{Kumamoto} et~al.}{2021}]{2021MNRAS.501.4490K}
{Kumamoto} H.,  et~al., 2021, \mn@doi [\mnras] {10.1093/mnras/staa3910}, \href {https://ui.adsabs.harvard.edu/abs/2021MNRAS.501.4490K} {501, 4490}

\bibitem[\protect\citeauthoryear{{Kumar}, {Lu}  \& {Bhattacharya}}{{Kumar} et~al.}{2017}]{2017MNRAS.468.2726K}
{Kumar} P.,  {Lu} W.,   {Bhattacharya} M.,  2017, \mn@doi [\mnras] {10.1093/mnras/stx665}, \href {https://ui.adsabs.harvard.edu/abs/2017MNRAS.468.2726K} {468, 2726}

\bibitem[\protect\citeauthoryear{{Lawson}, {Mayer}, {Osborne}  \& {Parkinson}}{{Lawson} et~al.}{1987}]{1987MNRAS.225..307L}
{Lawson} K.~D.,  {Mayer} C.~J.,  {Osborne} J.~L.,   {Parkinson} M.~L.,  1987, \mn@doi [\mnras] {10.1093/mnras/225.2.307}, \href {https://ui.adsabs.harvard.edu/abs/1987MNRAS.225..307L} {225, 307}

\bibitem[\protect\citeauthoryear{{Levin} et~al.,}{{Levin} et~al.}{2013}]{2013MNRAS.434.1387L}
{Levin} L.,  et~al., 2013, \mn@doi [\mnras] {10.1093/mnras/stt1103}, \href {https://ui.adsabs.harvard.edu/abs/2013MNRAS.434.1387L} {434, 1387}

\bibitem[\protect\citeauthoryear{{Lorimer}, {Bailes}, {McLaughlin}, {Narkevic}  \& {Crawford}}{{Lorimer} et~al.}{2007}]{2007Sci...318..777L}
{Lorimer} D.~R.,  {Bailes} M.,  {McLaughlin} M.~A.,  {Narkevic} D.~J.,   {Crawford} F.,  2007, \mn@doi [Science] {10.1126/science.1147532}, \href {https://ui.adsabs.harvard.edu/abs/2007Sci...318..777L} {318, 777}

\bibitem[\protect\citeauthoryear{{Macquart} \& {Johnston}}{{Macquart} \& {Johnston}}{2015}]{2015MNRAS.451.3278M}
{Macquart} J.-P.,  {Johnston} S.,  2015, \mn@doi [\mnras] {10.1093/mnras/stv1184}, \href {https://ui.adsabs.harvard.edu/abs/2015MNRAS.451.3278M} {451, 3278}

\bibitem[\protect\citeauthoryear{{Manchester} et~al.,}{{Manchester} et~al.}{2001}]{2001MNRAS.328...17M}
{Manchester} R.~N.,  et~al., 2001, \mn@doi [\mnras] {10.1046/j.1365-8711.2001.04751.x}, \href {https://ui.adsabs.harvard.edu/abs/2001MNRAS.328...17M} {328, 17}

\bibitem[\protect\citeauthoryear{{Manchester}, {Hobbs}, {Teoh}  \& {Hobbs}}{{Manchester} et~al.}{2005}]{2005AJ....129.1993M}
{Manchester} R.~N.,  {Hobbs} G.~B.,  {Teoh} A.,   {Hobbs} M.,  2005, \mn@doi [\aj] {10.1086/428488}, \href {https://ui.adsabs.harvard.edu/abs/2005AJ....129.1993M} {129, 1993}

\bibitem[\protect\citeauthoryear{{Miles}, {Shannon}, {Bailes}, {Reardon}, {Buchner}, {Middleton}  \& {Spiewak}}{{Miles} et~al.}{2022}]{2022MNRAS.510.5908M}
{Miles} M.~T.,  {Shannon} R.~M.,  {Bailes} M.,  {Reardon} D.~J.,  {Buchner} S.,  {Middleton} H.,   {Spiewak} R.,  2022, \mn@doi [\mnras] {10.1093/mnras/stab3549}, \href {https://ui.adsabs.harvard.edu/abs/2022MNRAS.510.5908M} {510, 5908}

\bibitem[\protect\citeauthoryear{{Miles} et~al.,}{{Miles} et~al.}{2023}]{2023MNRAS.519.3976M}
{Miles} M.~T.,  et~al., 2023, \mn@doi [\mnras] {10.1093/mnras/stac3644}, \href {https://ui.adsabs.harvard.edu/abs/2023MNRAS.519.3976M} {519, 3976}

\bibitem[\protect\citeauthoryear{{Morello} et~al.,}{{Morello} et~al.}{2019}]{2019MNRAS.483.3673M}
{Morello} V.,  et~al., 2019, \mn@doi [\mnras] {10.1093/mnras/sty3328}, \href {https://ui.adsabs.harvard.edu/abs/2019MNRAS.483.3673M} {483, 3673}

\bibitem[\protect\citeauthoryear{{Nathan}, {Miles}, {Ashton}, {Lasky}, {Thrane}, {Reardon}, {Shannon}  \& {Cameron}}{{Nathan} et~al.}{2023}]{2023arXiv230402793N}
{Nathan} R.~S.,  {Miles} M.~T.,  {Ashton} G.,  {Lasky} P.~D.,  {Thrane} E.,  {Reardon} D.~J.,  {Shannon} R.~M.,   {Cameron} A.~D.,  2023, \mn@doi [arXiv e-prints] {10.48550/arXiv.2304.02793}, \href {https://ui.adsabs.harvard.edu/abs/2023arXiv230402793N} {p. arXiv:2304.02793}

\bibitem[\protect\citeauthoryear{{Ord}, {van Straten}, {Hotan}  \& {Bailes}}{{Ord} et~al.}{2004}]{2004MNRAS.352..804O}
{Ord} S.~M.,  {van Straten} W.,  {Hotan} A.~W.,   {Bailes} M.,  2004, \mn@doi [\mnras] {10.1111/j.1365-2966.2004.07963.x}, \href {https://ui.adsabs.harvard.edu/abs/2004MNRAS.352..804O} {352, 804}

\bibitem[\protect\citeauthoryear{{Os{\l}owski}, {van Straten}, {Hobbs}, {Bailes}  \& {Demorest}}{{Os{\l}owski} et~al.}{2011}]{2011MNRAS.418.1258O}
{Os{\l}owski} S.,  {van Straten} W.,  {Hobbs} G.~B.,  {Bailes} M.,   {Demorest} P.,  2011, \mn@doi [\mnras] {10.1111/j.1365-2966.2011.19578.x}, \href {https://ui.adsabs.harvard.edu/abs/2011MNRAS.418.1258O} {418, 1258}

\bibitem[\protect\citeauthoryear{{Parthasarathy} et~al.,}{{Parthasarathy} et~al.}{2021}]{2021MNRAS.502..407P}
{Parthasarathy} A.,  et~al., 2021, \mn@doi [\mnras] {10.1093/mnras/stab037}, \href {https://ui.adsabs.harvard.edu/abs/2021MNRAS.502..407P} {502, 407}

\bibitem[\protect\citeauthoryear{{Petroff} et~al.,}{{Petroff} et~al.}{2014}]{2014ApJ...789L..26P}
{Petroff} E.,  et~al., 2014, \mn@doi [\apjl] {10.1088/2041-8205/789/2/L26}, \href {https://ui.adsabs.harvard.edu/abs/2014ApJ...789L..26P} {789, L26}

\bibitem[\protect\citeauthoryear{{Reardon} et~al.,}{{Reardon} et~al.}{2020}]{2020ApJ...904..104R}
{Reardon} D.~J.,  et~al., 2020, \mn@doi [\apj] {10.3847/1538-4357/abbd40}, \href {https://ui.adsabs.harvard.edu/abs/2020ApJ...904..104R} {904, 104}

\bibitem[\protect\citeauthoryear{{Rickett}}{{Rickett}}{1969}]{1969Natur.221..158R}
{Rickett} B.~J.,  1969, \mn@doi [\nat] {10.1038/221158a0}, \href {https://ui.adsabs.harvard.edu/abs/1969Natur.221..158R} {221, 158}

\bibitem[\protect\citeauthoryear{{Rickett}}{{Rickett}}{2001}]{2001Ap&SS.278....5R}
{Rickett} B.,  2001, \mn@doi [\apss] {10.1023/A:1013132101463}, \href {https://ui.adsabs.harvard.edu/abs/2001Ap&SS.278....5R} {278, 5}

\bibitem[\protect\citeauthoryear{{Rickett}, {Coles}  \& {Bourgois}}{{Rickett} et~al.}{1984}]{1984A&A...134..390R}
{Rickett} B.~J.,  {Coles} W.~A.,   {Bourgois} G.,  1984, \aap, \href {https://ui.adsabs.harvard.edu/abs/1984A&A...134..390R} {134, 390}

\bibitem[\protect\citeauthoryear{{Romani}, {Narayan}  \& {Blandford}}{{Romani} et~al.}{1986}]{1986MNRAS.220...19R}
{Romani} R.~W.,  {Narayan} R.,   {Blandford} R.,  1986, \mn@doi [\mnras] {10.1093/mnras/220.1.19}, \href {https://ui.adsabs.harvard.edu/abs/1986MNRAS.220...19R} {220, 19}

\bibitem[\protect\citeauthoryear{{Scheuer}}{{Scheuer}}{1968}]{1968Natur.218..920S}
{Scheuer} P.~A.~G.,  1968, \mn@doi [\nat] {10.1038/218920a0}, \href {https://ui.adsabs.harvard.edu/abs/1968Natur.218..920S} {218, 920}

\bibitem[\protect\citeauthoryear{{Shannon} et~al.,}{{Shannon} et~al.}{2014}]{2014MNRAS.443.1463S}
{Shannon} R.~M.,  et~al., 2014, \mn@doi [\mnras] {10.1093/mnras/stu1213}, \href {https://ui.adsabs.harvard.edu/abs/2014MNRAS.443.1463S} {443, 1463}

\bibitem[\protect\citeauthoryear{{Shin} et~al.,}{{Shin} et~al.}{2023}]{2023ApJ...944..105S}
{Shin} K.,  et~al., 2023, \mn@doi [\apj] {10.3847/1538-4357/acaf06}, \href {https://ui.adsabs.harvard.edu/abs/2023ApJ...944..105S} {944, 105}

\bibitem[\protect\citeauthoryear{{Siemens}, {Ellis}, {Jenet}  \& {Romano}}{{Siemens} et~al.}{2013}]{2013CQGra..30v4015S}
{Siemens} X.,  {Ellis} J.,  {Jenet} F.,   {Romano} J.~D.,  2013, \mn@doi [Classical and Quantum Gravity] {10.1088/0264-9381/30/22/224015}, \href {https://ui.adsabs.harvard.edu/abs/2013CQGra..30v4015S} {30, 224015}

\bibitem[\protect\citeauthoryear{{Smits}, {Kramer}, {Stappers}, {Lorimer}, {Cordes}  \& {Faulkner}}{{Smits} et~al.}{2009}]{2009A&A...493.1161S}
{Smits} R.,  {Kramer} M.,  {Stappers} B.,  {Lorimer} D.~R.,  {Cordes} J.,   {Faulkner} A.,  2009, \mn@doi [\aap] {10.1051/0004-6361:200810383}, \href {https://ui.adsabs.harvard.edu/abs/2009A&A...493.1161S} {493, 1161}

\bibitem[\protect\citeauthoryear{{Sobey} et~al.,}{{Sobey} et~al.}{2021}]{2021MNRAS.504..228S}
{Sobey} C.,  et~al., 2021, \mn@doi [\mnras] {10.1093/mnras/stab861}, \href {https://ui.adsabs.harvard.edu/abs/2021MNRAS.504..228S} {504, 228}

\bibitem[\protect\citeauthoryear{{Spiewak} et~al.,}{{Spiewak} et~al.}{2022}]{2022PASA...39...27S}
{Spiewak} R.,  et~al., 2022, \mn@doi [\pasa] {10.1017/pasa.2022.19}, \href {https://ui.adsabs.harvard.edu/abs/2022PASA...39...27S} {39, e027}

\bibitem[\protect\citeauthoryear{{Stairs}, {Thorsett}  \& {Camilo}}{{Stairs} et~al.}{1999}]{1999ApJS..123..627S}
{Stairs} I.~H.,  {Thorsett} S.~E.,   {Camilo} F.,  1999, \mn@doi [\apjs] {10.1086/313245}, \href {https://ui.adsabs.harvard.edu/abs/1999ApJS..123..627S} {123, 627}

\bibitem[\protect\citeauthoryear{{Stappers}}{{Stappers}}{2016}]{2016mks..confE..10S}
{Stappers} B.,  2016, in MeerKAT Science: On the Pathway to the SKA. p.~10, \mn@doi{10.22323/1.277.0010}

\bibitem[\protect\citeauthoryear{{Stappers} \& {Kramer}}{{Stappers} \& {Kramer}}{2016}]{2016mks..confE...9S}
{Stappers} B.,  {Kramer} M.,  2016, in MeerKAT Science: On the Pathway to the SKA. p.~9, \mn@doi{10.22323/1.277.0009}

\bibitem[\protect\citeauthoryear{{Stinebring} \& {Condon}}{{Stinebring} \& {Condon}}{1990}]{1990ApJ...352..207S}
{Stinebring} D.~R.,  {Condon} J.~J.,  1990, \mn@doi [\apj] {10.1086/168528}, \href {https://ui.adsabs.harvard.edu/abs/1990ApJ...352..207S} {352, 207}

\bibitem[\protect\citeauthoryear{{Stinebring}, {Smirnova}, {Hankins}, {Hovis}, {Kaspi}, {Kempner}, {Myers}  \& {Nice}}{{Stinebring} et~al.}{2000}]{2000ApJ...539..300S}
{Stinebring} D.~R.,  {Smirnova} T.~V.,  {Hankins} T.~H.,  {Hovis} J.~S.,  {Kaspi} V.~M.,  {Kempner} J.~C.,  {Myers} E.,   {Nice} D.~J.,  2000, \mn@doi [\apj] {10.1086/309201}, \href {https://ui.adsabs.harvard.edu/abs/2000ApJ...539..300S} {539, 300}

\bibitem[\protect\citeauthoryear{{Thornton} et~al.,}{{Thornton} et~al.}{2013}]{2013Sci...341...53T}
{Thornton} D.,  et~al., 2013, \mn@doi [Science] {10.1126/science.1236789}, \href {https://ui.adsabs.harvard.edu/abs/2013Sci...341...53T} {341, 53}

\bibitem[\protect\citeauthoryear{{Thorsett} \& {Chakrabarty}}{{Thorsett} \& {Chakrabarty}}{1999}]{1999ApJ...512..288T}
{Thorsett} S.~E.,  {Chakrabarty} D.,  1999, \mn@doi [\apj] {10.1086/306742}, \href {https://ui.adsabs.harvard.edu/abs/1999ApJ...512..288T} {512, 288}

\bibitem[\protect\citeauthoryear{{Toscano}, {Bailes}, {Manchester}  \& {Sandhu}}{{Toscano} et~al.}{1998}]{1998ApJ...506..863T}
{Toscano} M.,  {Bailes} M.,  {Manchester} R.~N.,   {Sandhu} J.~S.,  1998, \mn@doi [\apj] {10.1086/306282}, \href {https://ui.adsabs.harvard.edu/abs/1998ApJ...506..863T} {506, 863}

\bibitem[\protect\citeauthoryear{{Wang}, {Wang}, {Wang}, {Dai}  \& {Xie}}{{Wang} et~al.}{2023}]{2023MNRAS.tmp..221W}
{Wang} Z.,  {Wang} J.,  {Wang} N.,  {Dai} S.,   {Xie} J.,  2023, \mn@doi [\mnras] {10.1093/mnras/stad199}, \href {https://ui.adsabs.harvard.edu/abs/2023MNRAS.tmp..221W} {}

\bibitem[\protect\citeauthoryear{{Xue} et~al.,}{{Xue} et~al.}{2017}]{2017PASA...34...70X}
{Xue} M.,  et~al., 2017, \mn@doi [\pasa] {10.1017/pasa.2017.66}, \href {https://ui.adsabs.harvard.edu/abs/2017PASA...34...70X} {34, e070}

\bibitem[\protect\citeauthoryear{{You} et~al.,}{{You} et~al.}{2007}]{2007MNRAS.378..493Y}
{You} X.~P.,  et~al., 2007, \mn@doi [\mnras] {10.1111/j.1365-2966.2007.11617.x}, \href {https://ui.adsabs.harvard.edu/abs/2007MNRAS.378..493Y} {378, 493}

\bibitem[\protect\citeauthoryear{{Zic} et~al.,}{{Zic} et~al.}{2023}]{2023arXiv230616230Z}
{Zic} A.,  et~al., 2023, \mn@doi [arXiv e-prints] {10.48550/arXiv.2306.16230}, \href {https://ui.adsabs.harvard.edu/abs/2023arXiv230616230Z} {p. arXiv:2306.16230}

\makeatother
\end{thebibliography}




\appendix

\section{Flux calibration equation derivation}

In a folded pulsar observation, if there is negligible radio frequency interference, then the noise $N'$ in each
pulsar phase bin for $N_{\rm bin}$ bins, is just 

\begin{equation}
    N' = {{T_{\rm rec} + T_{\rm sky} \over{G \sqrt{BN_{\rm p}t/N_{\rm bin}}}}= N \sqrt{N_{\rm bin}}}
\end{equation}

In an RFI-free pulsar observation the mean flux density can be computed by integrating
the counts of the on-pulse bins and subtracting off the 
best-fit baseline, 
then determining the average flux density over the entire pulse period. 

The rms of the off-pulse region can be
equated to the system noise and sky temperature.
The baseline has an uncertainty $\sigma_{\rm b}$ determined
from the radiometer equation and the number of off-pulse bins in it which contributes to the uncertainty in the flux density

\begin{equation}
 \sigma_{\rm b} = {{N'}\over{\sqrt{N_{\rm off}}}} = {{N'}\over{\sqrt{N_{\rm bin}-N_{\rm on}}}}
\end{equation}

\noindent 
and similarly the uncertainty in the average height $h$ of the on-pulse bins has an uncertainty $\sigma_{\rm h}$ 

\begin{equation}
 \sigma_{\rm h} = {{N'}\over{\sqrt{N_{\rm on}}}}
\end{equation}

\noindent
So the area of the on-pulse region $A$ is just 

\begin{equation}
A=(h-b) N_{\rm on} = S {{N_{\rm bin}}\over{N_{\rm on}}} N_{\rm on} = S N_{\rm bin}
\end{equation}

\noindent
and the signal-to-noise ratio of the on-pulse region is
\begin{equation}
    snr=A/(\sigma_{\rm A}).
\end{equation}

\noindent
Now the uncertainty in $A$ is the quadrature sum of the uncertainties in
$h$ and $b$ multiplied by $N_{\rm on}$. So
\begin{equation}
    \sigma_{\rm A} = N_{\rm on} \sqrt{(\sigma_{\rm b})^2 + (\sigma_{\rm h})^2} = N' N_{\rm on} \sqrt{1/{(N_{\rm bin}- N_{\rm on})}
    + 1/(N_{\rm on})}
\end{equation}
\begin{equation}
    \sigma_{\rm A} = N' N_{\rm on} \sqrt{{{N_{\rm bin}}\over{(N_{\rm bin}-N_{\rm on})N_{\rm on}}}} = N' \sqrt{N_{\rm bin}} \sqrt{{N_{\rm on}}\over{N_{\rm bin}-N_{\rm on}}}
\end{equation}
\noindent so the $snr$ of the on-pulse region is just $A/\sigma_{\rm A}$
\begin{equation}
    snr = {{SN_{\rm bin}}\over{N' \sqrt{N_{\rm bin}} \sqrt{{{N_{\rm on}\over{N_{\rm bin}-N_{\rm on}}}}}}}= 
    {{S\sqrt{N_{\rm bin}}}\over{N'}} \sqrt{{{N_{\rm bin}-N_{\rm on}}\over{N_{\rm on}}}} 
\end{equation}

\begin{equation}
    snr = {S\over{N}}\sqrt{{N_{\rm bin}-N_{\rm on}}\over{N_{\rm on}}}
\end{equation}

\noindent so we can rearrange to get
\begin{equation}
    S = snr N \sqrt{{{N_{\rm on}}\over{N_{\rm bin}-N_{\rm on}}}} = snr {{T_{\rm rec} + T_{\rm sky}}\over{G \sqrt{BN_{\rm p}t}}} \sqrt{{{N_{\rm on}}\over{N_{\rm bin}-N_{\rm on}}}}
\end{equation}

where here $S$ is the mean flux density above the baseline.


\bsp	
\label{lastpage}
\end{document}